\documentclass[11pt, fullpage]{article}

\usepackage{fullpage}
\usepackage{setspace}
\usepackage{authblk}
\usepackage{cite}
\usepackage[colorlinks,citecolor=blue]{hyperref}
\usepackage{amssymb}
\usepackage{graphicx}
\usepackage{graphics}
\usepackage[cmex10]{amsmath}
\usepackage{latexsym,amsfonts}
\usepackage{amsthm}
\usepackage{subfigure}
\usepackage{enumitem}

\begin{document}
\title{On the Spurious Interior Resonance Modes of Time Domain Integral Equations for Analyzing Acoustic Scattering from Penetrable Objects} 

\author[1]{Rui Chen}
\author[2]{Yifei Shi} 
\author[3]{Sadeed Bin Sayed} 
\author[4]{Mingyu Lu} 
\author[1]{Hakan Bagci} 

\affil[1]{Division of Computer, Electrical, and Mathematical Science and Engineering, King Abdullah University of Science and Technology, Thuwal 23955-6900, Saudi Arabia \newline rui.chen@kaust.edu.sa}
\affil[2]{Department of Electronic Engineering, Jiangsu University of Technology, Changzhou, Jiangsu 213001, China}
\affil[3]{Halliburton Far East Pte. Ltd., 639940, Singapore}
\affil[4]{Department of Electrical and Computer Engineering, West Virginia University Institute of Technology, Beckley, WV 25801, United States}
\date{}
\renewcommand\Affilfont{\itshape}

\maketitle
\onehalfspacing

\begin{abstract}
\normalsize
The interior resonance problem of time domain integral equations (TDIEs) formulated to analyze acoustic field interactions on penetrable objects is investigated. Two types of TDIEs are considered: The first equation, which is termed the time domain potential integral equation (TDPIE) (in unknowns velocity potential and its normal derivative), suffers from the interior resonance problem, i.e., its solution is replete with spurious modes that are excited at the resonance frequencies of the acoustic cavity in the shape of the scatterer. Numerical experiments demonstrate that, unlike the frequency-domain integral equations, the amplitude of these modes in the time domain could be suppressed to a level that does not significantly affect the solution. The second equation is obtained by linearly combining TDPIE with its normal derivative. Weights of the combination are carefully selected to enable the numerical computation of the singular integrals. The solution of this equation, which is termed the time domain combined potential integral equation (TDCPIE), does not involve any spurious interior resonance modes.

\end{abstract}

\newpage
\doublespacing

\section{Introduction}\label{Introduction}
Many applications in engineering and physical sciences call for simulations of acoustic scattering from penetrable objects, i.e., scatterers that internally support nonzero velocity potential and pressure field~\cite{gonzalez2020boundary,pessoa2020acoustic,prospathopoulos2009underwater,bobrovnitskii2006new,kittappa1975acoustic,thomas2009balloon,kleinman1988,costabel1985direct,kress1978transmission,wu2012fast}. An acoustic scattering problem involving penetrable objects is also known as an acoustic transmission problem. The time-harmonic (frequency-domain) acoustic transmission problem can be analyzed by solving a set of integral equations enforced on the surface of the scatterer~\cite{colton2013integral}. This set of equations is obtained by using the Kirchhoff-Helmholtz theorem to express the scattered fields of the exterior and interior problems in terms of (unknown) velocity potential on the surface of the scatterer and its normal derivative. The exterior and interior problems involve the unbounded domains with the material properties (density and wave speed) of the background medium and the scatterer, respectively. Numerical schemes developed to solve these integral equations discretize only the surface of the scatterer and implicitly enforce the radiation condition at infinity~\cite{colton2013integral}, offering advantages over finite element and finite difference methods that directly solve the Helmholtz equation, and require a volumetric discretization of the whole computation domain and use absorbing boundary conditions on its surface to approximate the radiation condition. 

On the other hand, traditional integral equation formulations suffer from so-called ``interior resonance’’ problem~\cite{schenck1968,burton1971,zheng2015,schulz2020spurious,buffa2005regularized,boubendir2015regularized,jones1974integral,hwang1991retracted,qian2004non}. This problem is observed when the excitation frequency approaches any one of the resonance frequencies of the acoustic cavity in the shape of the scatterer and has the density and the wave speed of the background medium. At these frequencies the surface integral operator has a null space and the corresponding equation does not have a unique solution. Several approaches have been proposed to address the interior resonance problem of the frequency-domain integral equations. Examples of these include the combined Helmholtz integral equation formulation~\cite{schenck1968} and the Burton-Miller scheme~\cite{burton1971}. 

Even though interior resonance problem is well-studied for the frequency-domain integral equations, there are only a couple of studies that investigate the spurious interior resonance modes in the solution of time domain integral equations (TDIEs) of acoustics~\cite{ergin1999analysis,chappell2006stable,jang2013stabilization,kao2021time}. In~\cite{ergin1999analysis}, a spurious resonance-free Burton-Miller-type time domain combined field integral equation is formulated to analyze acoustic scattering from sound-rigid bodies. In~\cite{shi2013mode}, interior resonance modes observed in the solution of the time domain electric field integral equation (of electromagnetics) that is enforced on perfect electrically conducting scatterers are investigated. Theoretically, TDIEs should not admit any interior resonance modes since their solution is obtained under zero initial condition and the interior resonance modes do not satisfy this initial condition~\cite{shi2013mode, shanker2000}. But the interior resonance modes are still observed in the time domain solutions. It is discussed in~\cite{shi2013mode} that this is because of the numerical errors introduced due to discretization and matrix inversions carried out during time marching.

In this work, the interior resonance problem of two different TDIEs formulated to analyze the transient acoustic transmission problem is investigated: The first equation, which is termed time domain potential integral equation (TDPIE) (in unknowns velocity potential and its normal derivative) here, is the time-domain equivalent of the frequency-domain integral equation that is traditionally used in the literature to solve the acoustic transmission problem~\cite{kittappa1975acoustic,costabel1985direct,kress1978transmission,shanker2014quasi}. This equation suffers from the interior resonance problem. Its solution is replete with spurious modes that oscillate (without any decay) with the resonance frequencies of the acoustic cavity in the shape of the scatterer and has the density and the wave speed of the background medium~\cite{zheng2015}. These modes are excited when their resonance frequency is within the band of the excitation. In this work, it is demonstrated that unlike the frequency-domain integral equations, the amplitude of these modes in the time domain could be suppressed to a level, which does not significantly affect the solution, by increasing the accuracy of the discretization. This is achieved by using band-limited temporal basis functions and using smaller time step sizes. On the other hand, the second equation investigated in this work, which is termed time domain combined potential integral equation (TDCPIE), completely eliminates the interior resonance problem. The frequency-domain counterpart of TDCPIE has been introduced in~\cite{kleinman1988} and it has been theoretically shown that its frequency-domain solution is unique at the interior resonance frequencies. Numerical results in this work verify that TDCPIE does not admit any interior resonance modes. It should be noted here that TDCPIE is obtained by linearly combining TDPIE with its normal derivative. Coupling parameters of this combination are carefully selected to enable the computation of the singular integrals that appear in the expressions of the matrix elements resulting from the Nyström discretization in space~\cite{kang2001novel,rui2019jasa,rui2021tap,rui2019aces,rui2020aps,rui2021phd,rui2018aps,rui2019aps,rui2018apsb,abduljabbar2019extreme,noha2020solving,rui2020ursigassb,noha2017}.

The remainder of this paper is organized as follows. In Section~\ref{Formulation}, TDPIE and TDCPIE are derived. Section~\ref{NS} describes the spatial and temporal discretization schemes and the marching-on-in-time method that is used to solve the resulting matrix system. In Section~\ref{NR}, numerical results are presented to validate the accuracy of TDPIE and TDCPIE solutions and demonstrate the relationship between the numerical errors and the interior resonance modes observed in the solution of the TDPIE. Section~\ref{Conclusion} concludes the paper with a short summary.

\section{Formulation}\label{Formulation}
\begin{figure}
\centering
{\includegraphics[width=0.68\textwidth]{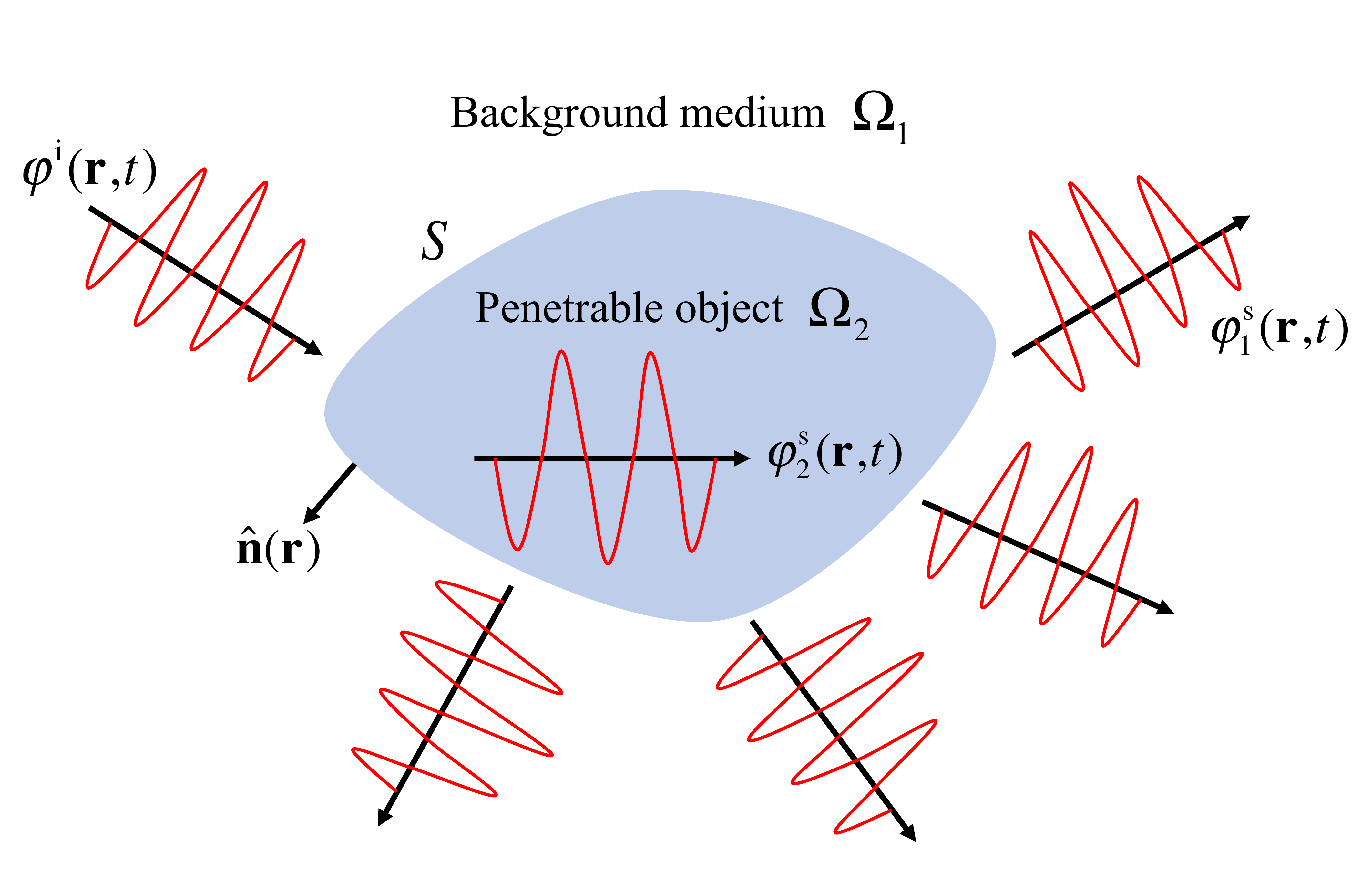}}\\
\caption{Description of the acoustic scattering problem.}
\label{fig1}
\end{figure}

Let ${\Omega _2}$ denote the support of an acoustically penetrable scatterer, which resides in an unbounded homogeneous background medium that is denoted by ${\Omega _1}$ (Fig.~\ref{fig1}). Let $S$ represent the surface that separates these two domains, i.e., the surface of the scatterer.  The wave speed and the density in ${\Omega _k}$, $k \in \{ 1,2\} $, are ${c_k}$ and ${\rho _k}$, respectively. An acoustic field with velocity potential ${\varphi ^{\mathrm{i}}}({\bf{r}},t)$ is incident on $S$. It is assumed that ${\varphi ^{\mathrm{i}}}({\bf{r}},t)$ is band-limited to maximum frequency ${f_{\max }}$ and vanishingly small for $t \le 0$ on ${\bf{r}} \in S$. In response to this excitation, scattered fields with velocity potentials $\varphi _k^{\mathrm{s}}({\bf{r}},t)$ are generated in ${\Omega _k}$. Total velocity potentials in  ${\Omega _1}$ and ${\Omega _2}$ are expressed as ${\varphi _1}({\bf{r}},t) = {\varphi ^{\mathrm{i}}}({\bf{r}},t) + \varphi _1^{\mathrm{s}}({\bf{r}},t)$ and ${\varphi _2}({\bf{r}},t) = \varphi _2^{\mathrm{s}}({\bf{r}},t)$, respectively. 
 
Using the Kirchhoff-Helmholtz theorem~\cite{pierce1981acoustics}, ${\partial _t}{\varphi _1}({\bf{r}},t)$ and ${\partial _t}{\varphi _2}({\bf{r}},t)$ are expressed as~\cite{shanker2014quasi}
\begin{align}
{\partial _t}{\varphi _1}({\bf{r}},t) &= {\partial _t}{\varphi ^{\mathrm{i}}}({\bf{r}},t) - {\partial _t}{S_1}[{\partial _n}{\varphi _1}]({\bf{r}},t) + {\partial _t}{D_1}[{\varphi _1}]({\bf{r}},t), {\bf{r}} \in {\Omega _1},\label{eq1}\\
{\partial _t}{\varphi _2}({\bf{r}},t) &= {\partial _t}{S_2}[{\partial _n}{\varphi _2}]({\bf{r}},t) - {\partial _t}{D_2}[{\varphi _2}]({\bf{r}},t),{\bf{r}} \in {\Omega _2}. \label{eq2}
\end{align}
Here, ${\partial _t}$ denotes the temporal derivative, ${\partial _n} = {\bf{\hat n}}({\bf{r}}) \cdot \nabla $, ${\bf{\hat n}}({\bf{r}})$ is the outward pointing unit normal at point ${\bf{r}} \in S$, and the spatio-temporal integral operators ${S_k}[x]({\bf{r}},t)$ and ${D_k}[x]({\bf{r}},t)$ are given by
\begin{align}
\nonumber {S_k}[x]({\bf{r}},t) &= \int_S {{G_k}(\left| {{\bf{r}} - {\bf{r'}}} \right|,t) * x({\bf{r'}},t)ds'},\\
\nonumber {D_k}[x]({\bf{r}},t) &= \int_S {{\partial _{n'}}{G_k}(\left| {{\bf{r}} - {\bf{r'}}} \right|,t) * x({\bf{r'}},t)ds'},
\end{align}
where ``$*$'' denotes temporal convolution and 
\begin{equation}
\nonumber {G_k}(\left| {{\bf{r}} - {\bf{r'}}} \right|,t) = \frac{\delta (t - \left| {{\bf{r}} - {\bf{r'}}} \right|/{c_k})}{4\pi \left| {{\bf{r}} - {\bf{r'}}} \right|}
\end{equation}
is the time domain Green function. Note that since the Green function is in the form of a Dirac delta function $\delta (t - {t_0})$, the temporal convolutions in operators ${S_k}[x]({\bf{r}},t)$ and ${D_k}[x]({\bf{r}},t)$ reduce retarded time integrals. 

On $S$, the acoustic pressure field and the normal component of the velocity field are continuous, i.e., the velocity potential satisfies the following boundary conditions~\cite{shanker2014quasi}:
\begin{align}
{\rho _1}{\partial _t}{\varphi _1}({\bf{r}},t) &= {\rho _2}{\partial _t}{\varphi _2}({\bf{r}},t), {\bf{r}} \in S,\label{eq3}\\
{\partial _n}{\varphi _1}({\bf{r}},t) &= {\partial _n}{\varphi _2}({\bf{r}},t), {\bf{r}} \in S. \label{eq4}
\end{align}

\subsection{TDPIE}\label{TDPIE}
Taking the limit of Eqs.~\eqref{eq1} and~\eqref{eq2} as ${\bf{r}}$ approaches $S$ from ${\Omega _k}$, $k \in \{1,2\}$ and inserting Eqs.~\eqref{eq3} and~\eqref{eq4} into the resulting equations yield TDPIE as~\cite{shanker2014quasi}
\begin{align}
\frac{1}{2}{\partial _t}{\varphi _1}({\bf{r}},t) &= {\partial _t}{\varphi ^{\rm{i}}}({\bf{r}},t) - {\partial _t}{S_1}[{\partial _n}{\varphi _1}]({\bf{r}},t) + {\partial _t}{\tilde D_1}[{\varphi _1}]({\bf{r}},t), {\bf{r}} \in S,\label{eq5}\\
\frac{{{\rho _1}}}{{2{\rho _2}}}{\partial _t}{\varphi _1}({\bf{r}},t) &= {\partial _t}{S_2}[{\partial _n}{\varphi _1}]({\bf{r}},t) - \frac{{{\rho _1}}}{{{\rho _2}}}{\partial _t}{\tilde D_2}[{\varphi _1}]({\bf{r}},t), {\bf{r}} \in S. \label{eq6}
\end{align}
Here, ``$\sim$'' on top of ${\tilde D_k}$ means that the space integral in ${D_k}$ is evaluated in the principal value sense~\cite{ishimaru1990}. 

\subsection{TDCPIE}\label{TDCPIE}
Taking the normal derivative of Eqs.~\eqref{eq1} and~\eqref{eq2} yields 
\begin{align}
{\partial _t}{\partial _n}{\varphi _1}({\bf{r}},t) &= {\partial _t}{\partial _n}{\varphi ^{\mathrm{i}}}({\bf{r}},t) - {\partial _t}{D'_1}[{\partial _n}{\varphi _1}]({\bf{r}},t) + {\partial _t}{N_1}[{\varphi _1}]({\bf{r}},t), {\bf{r}} \in {\Omega _1},\label{eq7}\\
{\partial _t}{\partial _n}{\varphi _2}({\bf{r}},t) &= {\partial _t}{D'_2}[{\partial _n}{\varphi _2}]({\bf{r}},t) - {\partial _t}{N_2}[{\varphi _2}]({\bf{r}},t), {\bf{r}} \in {\Omega _2}, 
\label{eq8}
\end{align}
where the spatio-temporal integral operators ${D'_k}[x]({\bf{r}},t)$ and ${N_k}[x]({\bf{r}},t)$ are given by
\begin{align}
\nonumber {D'_k}[x]({\bf{r}},t) &= \int_S {{\partial _n}{G_k}(\left| {{\bf{r}} - {\bf{r'}}} \right|,t) * x({\bf{r'}},t)ds'},\\
\nonumber {N_k}[x]({\bf{r}},t) &= \int_S {\partial _{nn'}^2{G_k}(\left| {{\bf{r}} - {\bf{r'}}} \right|,t) * x({\bf{r'}},t)ds'},
\end{align}
and $\partial _{nn'}^2 = {\partial _n}{\partial _{n'}}$ denotes the double normal derivative. Taking the limit of Eqs.~\eqref{eq7} and~\eqref{eq8} as ${\bf{r}}$ approaches $S$ from ${\Omega _k}$, $k \in \{ 1,2\} $ and inserting Eqs.~\eqref{eq3} and~\eqref{eq4} into the resulting equations yield the normal derivative of TDPIE in Eqs.~\eqref{eq5} and~\eqref{eq6} as
\begin{align}
\frac{1}{2}{\partial _t}{\partial _n}{\varphi _1}({\bf{r}},t) &= {\partial _t}{\partial _n}{\varphi ^{\mathrm{i}}}({\bf{r}},t) - {\partial _t}{\tilde D'_1}[{\partial _n}{\varphi _1}]({\bf{r}},t) + {\partial _t}{N_1}[{\varphi _1}]({\bf{r}},t), {\bf{r}} \in S,\label{eq9}\\
\frac{1}{2}{\partial _t}{\partial _n}{\varphi _1}({\bf{r}},t) &= {\partial _t}{\tilde D'_2}[{\partial _n}{\varphi _1}]({\bf{r}},t) - \frac{{{\rho _1}}}{{{\rho _2}}}{\partial _t}{N_2}[{\varphi _1}]({\bf{r}},t), {\bf{r}} \in S. \label{eq10}
\end{align}
Linearly combining Eqs.~\eqref{eq5}-\eqref{eq6} and~\eqref{eq9}-\eqref{eq10} as $\alpha_1$\eqref{eq5}+${\alpha _2}\frac{{{\rho _2}}}{{{\rho _1}}}$\eqref{eq6} and  ${\beta _1}$\eqref{eq9}+${\beta _2}$\eqref{eq10} yields TDCPIE as
\begin{align}
\nonumber &\frac{{{\alpha _1} + {\alpha _2}}}{2}{\partial _t}{\varphi _1}({\bf{r}},t) - {\alpha _1}{\partial _t}{\tilde D_1}[{\varphi _1}]({\bf{r}},t) + {\alpha _2}{\partial _t}{\tilde D_2}[{\varphi _1}]({\bf{r}},t) \\
&+ {\alpha _1}{\partial _t}{S_1}[{\partial _n}{\varphi _1}]({\bf{r}},t) - {\alpha _2}\frac{{{\rho _2}}}{{{\rho _1}}}{\partial _t}{S_2}[{\partial _n}{\varphi _1}]({\bf{r}},t) = {\alpha _1}{\partial _t}{\varphi ^{\mathrm{i}}}({\bf{r}},t), {\bf{r}} \in S,\label{eq11}\\
\nonumber &- {\beta _1}{\partial _t}{N_1}[{\varphi _1}]({\bf{r}},t) + {\beta _2}\frac{{{\rho _1}}}{{{\rho _2}}}{\partial _t}{N_2}[{\varphi _1}]({\bf{r}},t) + \frac{{{\beta _1} + {\beta _2}}}{2}{\partial _t}{\partial _n}{\varphi _1}({\bf{r}},t) \\
&+ {\beta _1}{\partial _t}{\tilde D'_1}[{\partial _n}{\varphi _1}]({\bf{r}},t) - {\beta _2}{\partial _t}{\tilde D'_2}[{\partial _n}{\varphi _1}]({\bf{r}},t) = {\beta _1}{\partial _t}{\partial _n}{\varphi ^{\mathrm{i}}}({\bf{r}},t), {\bf{r}} \in S. \label{eq12}
\end{align}
Here, ${\alpha _k}$ and ${\beta _k}$ are real constants.

\section{Numerical Solution}\label{NS}
To solve the coupled systems of equations Eqs.~\eqref{eq5}-\eqref{eq6} and~\eqref{eq11}-\eqref{eq12} numerically, first, $S$ is discretized into a mesh of curvilinear triangles and surface unknowns ${\varphi _1}({\bf{r}},t)$ and ${\partial _n}{\varphi _1}({\bf{r}},t)$ are expanded in space and time as
\begin{align}
{\varphi _1}({\bf{r}},t) &= \sum\limits_{i = 1}^{{N_{\mathrm{t}}}} {\sum\limits_{q = 1}^{{N_{\mathrm{p}}}} {\sum\limits_{n = 1}^{{N_{\mathrm{n}}}} {{{\left. {{\bf{I}}_i^1} \right|}_{qn}}\vartheta ({\bf{r}}){\ell _{qn}}({\bf{r}})T(t - i\Delta t)} } } ,\label{eq13}\\
{\partial _n}{\varphi _1}({\bf{r}},t) &= \sum\limits_{i = 1}^{{N_{\mathrm{t}}}} {\sum\limits_{q = 1}^{{N_{\mathrm{p}}}} {\sum\limits_{n = 1}^{{N_{\mathrm{n}}}} {{{\left. {{\bf{I}}_i^2} \right|}_{qn}}\vartheta ({\bf{r}}){\ell _{qn}}({\bf{r}})T(t - i\Delta t)} } }. \label{eq14}
\end{align}
In Eqs.~\eqref{eq13} and~\eqref{eq14}, ${N_{\mathrm{t}}}$ is the number of time steps, ${N_{\mathrm{p}}}$ is the number of curvilinear triangles, ${N_{\mathrm{n}}}$ is the number of interpolation nodes on each triangle, ${\ell _{qn}}({\bf{r}})$ is the Lagrange interpolation function defined at ${{\bf{r}}_{qn}}$ (node $n$ on triangle $q$)~\cite{kang2001novel}, $\vartheta ({\bf{r}})$ is the inverse of the Jacobian of the coordinate transformation between the unit right triangle and the Cartesian coordinate system, $T(t)$ is the temporal basis function which is constructed using the band-limited approximate prolate spherical wave (APSW) function ~\cite{knab1979}, $\Delta t$ is the time step size, and ${\left. {{\bf{I}}_i^1} \right|_{qn}}$ and ${\left. {{\bf{I}}_i^2} \right|_{qn}}$ are the unknown expansion coefficients to be solved for.

Substituting Eqs.~\eqref{eq13}-\eqref{eq14} into Eqs.~\eqref{eq5}-\eqref{eq6} and~\eqref{eq11}-\eqref{eq12} and point-testing the resulting equations in space at ${{\bf{r}}_{pm}}$, $p = 1,...,{N_{\mathrm{p}}}$, $m = 1,...,{N_{\mathrm{n}}}$ (i.e., Nystr\"{o}m discretization in space), and in time $t = j\Delta t$ yield the following system of matrix equations
\begin{align}
{{\bf{Z}}_0}{{\bf{I}}_j} = {{\bf{V}}_j} - \sum\limits_{i = 1}^{j - 1} {{{\bf{Z}}_{j - i}}{{\bf{I}}_i}}  - \sum\limits_{i = j + 1}^{j + {N_{{\mathrm{hw}}}}} {{{\bf{Z}}_{j - i}}{{\bf{I}}_i}}, j = 1,...,{N_{\mathrm{t}}}. \label{eq15}
\end{align}
Here, ${N_{{\mathrm{hw}}}}$ is the half-width of the APSW function $T(t)$, ${{\bf{I}}_j} = {[\begin{array}{*{20}{c}}
{{\bf{I}}_j^1}&{{\bf{I}}_j^2}
\end{array}]^T}$, where ${\bf{I}}_j^1$ and ${\bf{I}}_j^2$ store the unknown expansion coefficients in Eqs.~\eqref{eq13} and~\eqref{eq14},   ${{\bf{V}}_j} = {[\begin{array}{*{20}{c}}
{{\bf{V}}_j^1}&{{\bf{V}}_j^2}
\end{array}]^T}$ store the tested excitation vectors at time $t = j\Delta t$, and 
\begin{equation}
\nonumber
{{\bf{Z}}_{j - i}} = \left[ {\begin{array}{*{20}{c}}
{{\bf{Z}}_{j - i}^{11}}&{{\bf{Z}}_{j - i}^{12}}\\{{\bf{Z}}_{j - i}^{21}}&{{\bf{Z}}_{j - i}^{22}}
\end{array}} \right]
\end{equation}
store the discretized retarted time integrals between nodes of mesh elements. Expressions of elements of ${{\bf{V}}_j}$ and  ${{\bf{Z}}_{j - i}}$  for TDPIE and TDCPIE are provided in Sections~\ref{ETDPIE} and~\ref{ETDCPIE}, respectively. Note that the system of matrix equations in Eq.~\eqref{eq15} is not causal, i.e.,  ${{\bf{I}}_j}$ can not be solved for without knowing ``future'' unknowns ${{\bf{I}}_{j + 1}}$, ${{\bf{I}}_{j + 2}}$, …, ${{\bf{I}}_{j + {N_{{\mathrm{hw}}}}}}$ [see the second summation on the right-hand side of Eq.~\eqref{eq15}]. The extrapolation scheme developed in~\cite{sayed2015} is used here to express these future unknowns ${{\bf{I}}_{j + 1}}$, ${{\bf{I}}_{j + 2}}$, …, ${{\bf{I}}_{j + {N_{{\mathrm{hw}}}}}}$ in terms of ``past/current'' unknowns  ${{\bf{I}}_{j - N + 1}}$, …, ${{\bf{I}}_{j - 1}}$, ${{\bf{I}}_j}$, where $N$  is the number of samples used in the extrapolation. Inserting this expression into Eq.~\eqref{eq15} converts it into a causal form as
\begin{align}
{{\bf{\bar Z}}_0}{{\bf{I}}_j} = {{\bf{V}}_j} - \sum\limits_{i = 1}^{j - 1} {{{{\bf{\bar Z}}}_{j - i}}{{\bf{I}}_i}}, j = 1,...,{N_{\mathrm{t}}}. \label{eq16}
\end{align}
The modified matrices ${{\bf{\bar Z}}_{j - i}}$ in Eq.~\eqref{eq16} are obtained from ${{\bf{Z}}_{j - i}}$ in Eq.~\eqref{eq15} using the expressions given in~\cite{sayed2015}. The system of matrix equations in Eq.~\eqref{eq16} is now in a form that can be recursively solved for the unknown coefficient vectors ${{\bf{I}}_j}$, $j = 1,...,{N_{\mathrm{t}}}$ via time marching as briefly described next. For $j = 1$, ${{\bf{I}}_1}$ is found by solving Eq.~\eqref{eq16} with right-hand side ${{\bf{V}}_1}$ (contribution from the summation is zero at the first time step). For $j = 2$,  the right-hand side of Eq.~\eqref{eq16} is computed by subtracting ${{\bf{\bar Z}}_1}{{\bf{I}}_1}$ (only term coming from the summation) from ${{\bf{V}}_2}$. Then, ${{\bf{I}}_2}$ is found by solving Eq.~\eqref{eq16} with this right-hand side. For $j = 3, $ the right-hand side of Eq.~\eqref{eq16} is computed by subtracting ${{\bf{\bar Z}}_2}{{\bf{I}}_1} + {{\bf{\bar Z}}_1}{{\bf{I}}_2}$ from ${{\bf{V}}_3}$. Then, Eq.~\eqref{eq16} is solved for ${{\bf{I}}_3}$.  This recursive time marching algorithm is continued until all ${{\bf{I}}_j}$, $j = 1,...,{N_{\mathrm{t}}}$ are obtained. 

In the next two sections, Sections~\ref{ETDPIE} and~\ref{ETDCPIE}, the expressions of the elements of ${{\bf{V}}_j}$ and ${{\bf{Z}}_{j - i}}$ in Eq.~\eqref{eq15} are provided for TDPIE and TDCPIE, respectively.

\subsection{Elements of  ${{\bf{V}}_j}$ and ${{\bf{Z}}_{j - i}}$ for TDPIE}\label{ETDPIE}
The elements of  ${\bf{V}}_j^1$ and ${\bf{V}}_j^2$ are given as ${\left. {{\bf{V}}_j^1} \right|_{pm}} = {\left. {{\partial _t}{\varphi ^{\mathrm{i}}}({{\bf{r}}_{pm}},t)} \right|_{t = j\Delta t}}$ and ${\left. {{\bf{V}}_j^2} \right|_{pm}} = 0$, respectively.  The elements of ${{\bf{Z}}_{j - i}}$ are expressed as 
\begin{align}
\nonumber {\left. {{\bf{Z}}_{j - i}^{11}} \right|_{pm,qn}} &= \frac{1}{2}\vartheta ({{\bf{r}}_{pm}}){\left. {{\partial _t}T(t)} \right|_{t = (j - i)\Delta t}}{\delta _{pq}}{\delta _{mn}}\\
\nonumber & - {\left. {  {\partial _t}\int_{{S_q}} {{\partial _{n'}}{G_1}(R,t) * T(t - i\Delta t)\vartheta ({\bf{r'}}){\ell _{qn}}({\bf{r'}})ds'} } \right|_{t = j\Delta t}}, \\
\nonumber {\left. {{\bf{Z}}_{j - i}^{12}} \right|_{pm,qn}} &= {\left. {{\partial _t}\int_{{S_q}} {{G_1}(R,t) * T(t - i\Delta t)\vartheta ({\bf{r'}}){\ell _{qn}}({\bf{r'}})ds'} } \right|_{t = j\Delta t}},\\
\nonumber {\left. {{\bf{Z}}_{j - i}^{21}} \right|_{pm,qn}} &= \frac{1}{2}\vartheta ({{\bf{r}}_{pm}}){\left. {{\partial _t}T(t)} \right|_{t = (j - i)\Delta t}}{\delta _{pq}}{\delta _{mn}} \\
\nonumber &+ {\left. {{\partial _t}\int_{{S_q}} {{\partial _{n'}}{G_2}(R,t) * T(t - i\Delta t)\vartheta ({\bf{r'}}){\ell _{qn}}({\bf{r'}})ds'} } \right|_{t = j\Delta t}}, \\
{\left. {{\bf{Z}}_{j - i}^{22}} \right|_{pm,qn}} &=  - \frac{{{\rho _2}}}{{{\rho _1}}}{\left. {{\partial _t}\int_{{S_q}} {{G_2}(R,t) * T(t - i\Delta t)\vartheta ({\bf{r'}}){\ell _{qn}}({\bf{r'}})ds'} } \right|_{t = j\Delta t}}.\label{eq17}
\end{align}
Here, $R=\left| {{{\bf{r}}_{pm}} - {\bf{r'}}} \right|$, ${S_q}$ is the surface of the curvilinear triangle $q$, and ${\delta _{pq}} = 1$ for $p = q$, and ${\delta _{pq}} = 0$ for $p \ne q$.

\subsection{Elements of ${{\bf{V}}_j}$ and ${{\bf{Z}}_{j - i}}$ for TDCPIE}\label{ETDCPIE}
The elements of  ${\bf{V}}_j^1$ and ${\bf{V}}_j^2$ are given as ${\left. {{\bf{V}}_j^1} \right|_{pm}} = {\alpha _1}{\left. {{\partial _t}{\varphi ^{\mathrm{i}}}({{\bf{r}}_{pm}},t)} \right|_{t = j\Delta t}}$  and ${\left. {{\bf{V}}_j^2} \right|_{pm}} = {\beta _1}{\left. {{\partial _t}{\partial _n}{\varphi ^{\mathrm{i}}}({{\bf{r}}_{pm}},t)} \right|_{t = j\Delta t}}$, respectively. The elements of ${{\bf{Z}}_{j - i}}$ are expressed as 
\begin{align}
\nonumber {\left. {{\bf{Z}}_{j - i}^{11}} \right|_{pm,qn}} & = \frac{{{\alpha _1} + {\alpha _2}}}{2}\vartheta ({{\bf{r}}_{pm}}){\left. {{\partial _t}T(t)} \right|_{t = (j - i)\Delta t}}{\delta _{pq}}{\delta _{mn}}\\
\nonumber &- {\left. {{\alpha _1}{\partial _t}\int_{{S_q}} {{\partial _{n'}}{G_1}(R,t) * T(t - i\Delta t)\vartheta ({\bf{r'}}){\ell _{qn}}({\bf{r'}})ds'} } \right|_{t = j\Delta t}}  \\
\nonumber &+ {\left. {{\alpha _2}{\partial _t}\int_{{S_q}} {{\partial _{n'}}{G_2}(R,t) * T(t - i\Delta t)\vartheta ({\bf{r'}}){\ell _{qn}}({\bf{r'}})ds'} } \right|_{t = j\Delta t}}, \\
\nonumber {\left. {{\bf{Z}}_{j - i}^{12}} \right|_{pm,qn}} & = {\left. {{\alpha _1}{\partial _t}\int_{{S_q}} {{G_1}(R,t) * T(t - i\Delta t)\vartheta ({\bf{r'}}){\ell _{qn}}({\bf{r'}})ds'} } \right|_{t = j\Delta t}}  \\
\nonumber &- {\alpha _2}\frac{{{\rho _2}}}{{{\rho _1}}}{\left. {{\partial _t}\int_{{S_q}} {{G_2}(R,t) * T(t - i\Delta t)\vartheta ({\bf{r'}}){\ell _{qn}}({\bf{r'}})ds'} } \right|_{t = j\Delta t}},\\
\nonumber {\left. {{\bf{Z}}_{j - i}^{21}} \right|_{pm,qn}} &=  - {\beta _1}{\left. {{\partial _t}\int_{{S_q}} {\partial _{nn'}^2{G_1}(R,t) * T(t - i\Delta t)\vartheta ({\bf{r'}}){\ell _{qn}}({\bf{r'}})ds'} } \right|_{t = j\Delta t}}  \\
\nonumber &+ {\left. {\beta _2}{\frac{{{\rho _1}}}{{{\rho _2}}}{\partial _t}\int_{{S_q}} {\partial _{nn'}^2{G_2}(R,t) * T(t - i\Delta t)\vartheta ({\bf{r'}}){\ell _{qn}}({\bf{r'}})ds'} } \right|_{t = j\Delta t}},\\
\nonumber {\left. {{\bf{Z}}_{j - i}^{22}} \right|_{pm,qn}} & = \frac{{{\beta _1} + {\beta _2}}}{2}\vartheta ({{\bf{r}}_{pm}}){\left. {{\partial _t}T(t)} \right|_{t = (j - i)\Delta t}}{\delta _{pq}}{\delta _{mn}}\\
\nonumber &+ {\left. {{\beta _1}{\partial _t}\int_{{S_q}} {{\partial _n}{G_1}(R,t) * T(t - i\Delta t)\vartheta ({\bf{r'}}){\ell _{qn}}({\bf{r'}})ds'} } \right|_{t = j\Delta t}}  \\
&- {\beta _2}{\left. {{\partial _t}\int_{{S_q}} {{\partial _n}{G_2}(R,t) * T(t - i\Delta t)\vartheta ({\bf{r'}}){\ell _{qn}}({\bf{r'}})ds'} } \right|_{t = j\Delta t}}.\label{eq18}
\end{align}
Computation of the matrix elements in Eqs.~\eqref{eq17} and~\eqref{eq18} calls for treatment of the singularities of the spatial integrals. This singularity treatment is described in the next section, Section~\ref{CSI}. 

\subsection{Computation of Singular Integrals}\label{CSI}
There are four kinds of integrands in Eqs.~\eqref{eq17} and~\eqref{eq18}: ${G_k}(R,t) * T(t - i\Delta t)$, ${\partial _n}{G_k}(R,t) * T(t - i\Delta t)$, ${\partial _{n'}}{G_k}(R,t) * T(t - i\Delta t)$, and $\partial _{nn'}^2{G_k}(R,t) * T(t - i\Delta t)$. When $p = q$, and as ${\bf{r'}}$ approaches ${{\bf{r}}_{pm}}$, the integrals of ${G_k}(R,t) * T(t - i\Delta t)$ and ${\partial _n}{G_k}(R,t) * T(t - i\Delta t)$ become weakly-singular, the integral of ${\partial _{n'}}{G_k}(R,t) * T(t - i\Delta t)$ becomes strongly-singular, and the integral of $\partial _{nn'}^2{G_k}(R,t) * T(t - i\Delta t)$ becomes hyper-singular~\cite{liu1992weakly}. The weakly-singular integrals are computed using the Duffy transformation~\cite{duffy1982quadrature}. The strongly-singular integrals in Eq.~\eqref{eq17} are computed using the approach described in~\cite{rui2019jasa}. The strongly- and hyper-singular integrals in Eq.~\eqref{eq18} are computed using the method described as follows.

$T(t - R/{c_k})$ is expanded using the Taylor series around $R = 0$ as
\begin{align}
T(t - R/{c_k}) = \sum\limits_{u = 0}^\infty  {\frac{{{{( - 1)}^u}\partial _t^uT(t)}}{{u!c_k^u}}{R^u}}. \label{eq19}
\end{align}
Using this expansion in ${\partial _{n'}}\{{T(t - R/{c_k})}/(4\pi R)\}$ and $\partial _{nn'}^2\{T(t - R/{c_k})/(4\pi R)\}$ yields 
\begin{align}
\nonumber {\partial _{n'}}\frac{{T(t - R/{c_k})}}{{4\pi R}} &= \frac{{({\bf{\hat n'}} \cdot {\bf{\hat R}})}}{{4\pi }}\left[ {\frac{{{\partial _t}T(t - R/{c_k})}}{{{c_k}R}} + \frac{{T(t - R/{c_k})}}{{{R^2}}}} \right]\\
\nonumber &= \frac{{({\bf{\hat n'}} \cdot {\bf{\hat R}})}}{{4\pi }}\left[ {\sum\limits_{u = 0}^\infty  {\frac{{{{( - 1)}^u}\partial _t^{u + 1}T(t)}}{{u!c_k^{u + 1}}}{R^{u - 1}}}  + \sum\limits_{u = 0}^\infty  {\frac{{{{( - 1)}^u}\partial _t^uT(t)}}{{u!c_k^u}}{R^{u - 2}}} } \right]\\
\nonumber &= \frac{{({\bf{\hat n'}} \cdot {\bf{\hat R}})}}{{4\pi }}\left[ \frac{{{\partial _t}T(t)}}{{{c_k}}}{R^{ - 1}} + T(t){R^{ - 2}} + \frac{{( - 1){\partial _t}T(t)}}{{{c_k}}}{R^{ - 1}} + O({R^{u \ge 0}}) \right]\\
 &= ({\bf{\hat n'}} \cdot {\bf{\hat R}})\frac{{T(t)}}{{4\pi {R^2}}} + O({R^{u \ge 0}}), \label{eq20}
\end{align}
\begin{align}
\nonumber \partial _{nn'}^2\frac{{T(t - R/{c_k})}}{{4\pi R}}& = \frac{{({\bf{\hat n}} \cdot {\bf{\hat n'}})}}{{4\pi }}\left[ {\frac{{{\partial _t}T(t - R/{c_k})}}{{{c_k}{R^2}}} + \frac{{T(t - R/{c_k})}}{{{R^3}}}} \right] \\
\nonumber &- \frac{{({\bf{\hat n}} \cdot {\bf{\hat R}})({\bf{\hat n'}} \cdot {\bf{\hat R}})}}{{4\pi }}\left[ {\frac{{\partial _t^2T(t - R/{c_k})}}{{c_k^2R}} + \frac{{3{\partial _t}T(t - R/{c_k})}}{{{c_k}{R^2}}} + \frac{{3T(t - R/{c_k})}}{{{R^3}}}} \right]\\
\nonumber & = \frac{{({\bf{\hat n}} \cdot {\bf{\hat n'}})}}{{4\pi }}\left[ {\sum\limits_{u = 0}^\infty  {\frac{{{{( - 1)}^u}\partial _t^{u + 1}T(t)}}{{u!c_k^{u + 1}}}{R^{u - 2}}}  + \sum\limits_{u = 0}^\infty  {\frac{{{{( - 1)}^u}\partial _t^uT(t)}}{{u!c_k^u}}{R^{u - 3}}} } \right]\\
\nonumber & -\frac{{({\bf{\hat n}} \cdot {\bf{\hat R}})({\bf{\hat n'}} \cdot {\bf{\hat R}})}}{{4\pi }} \left[ \sum\limits_{u = 0}^\infty  {\frac{{{{( - 1)}^u}\partial _t^{u + 2}T(t)}}{{u!c_k^{u + 2}}}{R^{u - 1}}}  + 3\sum\limits_{u = 0}^\infty  {\frac{{{{( - 1)}^u}\partial _t^{u + 1}T(t)}}{{u!c_k^{u + 1}}}{R^{u - 2}}}\right. \\
\nonumber & \left. + 3\sum\limits_{u = 0}^\infty  {\frac{{{{( - 1)}^u}\partial _t^uT(t)}}{{u!c_k^u}}{R^{u - 3}}} \right]\\
\nonumber & = \frac{{({\bf{\hat n}} \cdot {\bf{\hat n'}})}}{{4\pi }}\left[ \frac{{{\partial _t}T(t)}}{{{c_k}}}{R^{ - 2}} + \frac{{( - 1)\partial _t^2T(t)}}{{c_k^2}}{R^{ - 1}} + T(t){R^{ - 3}} + \frac{{( - 1){\partial _t}T(t)}}{{{c_k}}}{R^{ - 2}}\right. \\
\nonumber &+ \left. \frac{{\partial _t^2T(t)}}{{2c_k^2}}{R^{ - 1}} + O({R^{u \ge 0}}) \right]\\
\nonumber &- \frac{{({\bf{\hat n}} \cdot {\bf{\hat R}})({\bf{\hat n'}} \cdot {\bf{\hat R}})}}{{4\pi }}\left[ \frac{{\partial _t^2T(t)}}{{c_k^2}}{R^{ - 1}} + 3\frac{{{\partial _t}T(t)}}{{{c_k}}}{R^{ - 2}} + 3\frac{{( - 1)\partial _t^2T(t)}}{{c_k^2}}{R^{ - 1}} \right. \\
\nonumber &+ \left. 3T(t){R^{ - 3}} + 3\frac{{( - 1){\partial _t}T(t)}}{{{c_k}}}{R^{ - 2}} + 3\frac{{\partial _t^2T(t)}}{{2c_k^2}}{R^{ - 1}} + O({R^{u \ge 0}}) \right]\\
\nonumber & = \left[ {({\bf{\hat n}} \cdot {\bf{\hat n'}}) - 3({\bf{\hat n}} \cdot {\bf{\hat R}})({\bf{\hat n'}} \cdot {\bf{\hat R}})} \right]\frac{{T(t)}}{{4\pi {R^3}}} \\
&+ \left[ { - ({\bf{\hat n}} \cdot {\bf{\hat n'}}) + ({\bf{\hat n}} \cdot {\bf{\hat R}})({\bf{\hat n'}} \cdot {\bf{\hat R}})} \right]\frac{{\partial _t^2T(t)}}{{8\pi c_k^2R}} + O({R^{u \ge 0}}). \label{eq21}
\end{align}
Here, ${\bf{\hat R}}=({{\bf{r}}_{pm}} - {\bf{r'}})/R$ and $O({R^{u \ge 0}})$ represents the higher-order terms that are not singular as $R \to 0$. The singular terms on the right hands of Eqs.~\eqref{eq20} and~\eqref{eq21} are subtracted from ${\partial _{n'}}{G_k}(R,t) * T(t - i\Delta t)$ and $\partial _{nn'}^2{G_k}(R,t) * T(t - i\Delta t)$, respectively. Then, the integrals of these terms are added back to yield the final expressions for ${\left. {{\bf{Z}}_{j - i}^{11}} \right|_{pm,qn}}$ and ${\left. {{\bf{Z}}_{j - i}^{21}} \right|_{pm,qn}}$ in~\eqref{eq18} as
\begin{align}
\nonumber {\left. {{\bf{Z}}_{j - i}^{11}} \right|_{pm,qn}} & =  - {\alpha _1}{\partial _t}{\left. {\int_{{S_q}} {\left[ {{\partial _{n'}}\frac{{T(t - R/{c_1})}}{{4\pi R}} - ({\bf{\hat n'}} \cdot {\bf{\hat R}})\frac{{T(t)}}{{4\pi {R^2}}}} \right]\vartheta ({\bf{r'}}){\ell _{qn}}({\bf{r'}})ds'} } \right|_{t = (j - i)\Delta t}}\\
\nonumber & - {\alpha _1}{\left. {\int_{{S_q}} {({\bf{\hat n'}} \cdot {\bf{\hat R}})\frac{{{\partial _t}T(t)}}{{4\pi {R^2}}}\vartheta ({\bf{r'}}){\ell _{qn}}({\bf{r'}})ds'} } \right|_{t = (j - i)\Delta t}}\\
\nonumber & + {\alpha _2}{\partial _t}{\left. {\int_{{S_q}} {\left[ {{\partial _{n'}}\frac{{T(t - R/{c_2})}}{{4\pi R}} - ({\bf{\hat n'}} \cdot {\bf{\hat R}})\frac{{T(t)}}{{4\pi {R^2}}}} \right]\vartheta ({\bf{r'}}){\ell _{qn}}({\bf{r'}})ds'} } \right|_{t = (j - i)\Delta t}}\\
 &+ {\alpha _2}{\left. {\int_{{S_q}} {({\bf{\hat n'}} \cdot {\bf{\hat R}})\frac{{{\partial _t}T(t)}}{{4\pi {R^2}}}\vartheta ({\bf{r'}}){\ell _{qn}}({\bf{r'}})ds'} } \right|_{t = (j - i)\Delta t}}, \label{eq22}
\end{align}
\begin{align}
\nonumber {\left. {{\bf{Z}}_{j - i}^{21}} \right|_{pm,qn}} & =  - {\beta _1}{\partial _t}\int_{{S_q}}\left\{\partial _{nn'}^2\frac{{T(t - R/{c_1})}}{{4\pi R}} - \left[ {({\bf{\hat n}} \cdot {\bf{\hat n'}}) - 3({\bf{\hat n}} \cdot {\bf{\hat R}})({\bf{\hat n'}} \cdot {\bf{\hat R}})} \right]\frac{{T(t)}}{{4\pi {R^3}}}\right.\\
\label{eq23}&-\left.\left.\left[ { - ({\bf{\hat n}} \cdot {\bf{\hat n'}}) + ({\bf{\hat n}} \cdot {\bf{\hat R}})({\bf{\hat n'}} \cdot {\bf{\hat R}})} \right]\frac{{\partial _t^2T(t)}}{{8\pi c_1^2R}}\right\}\vartheta ({\bf{r'}}){\ell _{qn}}({\bf{r'}})ds'\right|_{t= (j - i)\Delta t}  \\
\nonumber &{\left. { - {\beta _1}\int_{{S_q}} {\left[ {({\bf{\hat n}} \cdot {\bf{\hat n'}}) - 3({\bf{\hat n}} \cdot {\bf{\hat R}})({\bf{\hat n'}} \cdot {\bf{\hat R}})} \right]\frac{{{\partial _t}T(t)}}{{4\pi {R^3}}}\vartheta ({\bf{r'}}){\ell _{qn}}({\bf{r'}})ds'} } \right|_{t = (j - i)\Delta t}}\\
\nonumber &{\left. { - {\beta _1}\int_{{S_q}} {\left[ { - ({\bf{\hat n}} \cdot {\bf{\hat n'}}) + ({\bf{\hat n}} \cdot {\bf{\hat R}})({\bf{\hat n'}} \cdot {\bf{\hat R}})} \right]\frac{{\partial _t^3T(t)}}{{8\pi c_1^2R}}\vartheta ({\bf{r'}}){\ell _{qn}}({\bf{r'}})ds'} } \right|_{t = (j - i)\Delta t}}\\
\nonumber & + {\beta _2}\frac{{{\rho _1}}}{{{\rho _2}}}{\partial _t}\int_{{S_q}} \left\{\partial _{nn'}^2\frac{{T(t - R/{c_2})}}{{4\pi R}} - \left[ {({\bf{\hat n}} \cdot {\bf{\hat n'}}) - 3({\bf{\hat n}} \cdot {\bf{\hat R}})({\bf{\hat n'}} \cdot {\bf{\hat R}})} \right]\frac{{T(t)}}{{4\pi {R^3}}}\right. \\
\nonumber &- \left.\left.\left[ { - ({\bf{\hat n}} \cdot {\bf{\hat n'}}) + ({\bf{\hat n}} \cdot {\bf{\hat R}})({\bf{\hat n'}} \cdot {\bf{\hat R}})} \right]\frac{{\partial _t^2T(t)}}{{8\pi c_2^2R}} \right\}\vartheta ({\bf{r'}}){\ell _{qn}}({\bf{r'}})ds' \right|_{t = (j - i)\Delta t}\\
\nonumber & + {\left. {{\beta _2}\frac{{{\rho _1}}}{{{\rho _2}}}\int_{{S_q}} {\left[ {({\bf{\hat n}} \cdot {\bf{\hat n'}}) - 3({\bf{\hat n}} \cdot {\bf{\hat R}})({\bf{\hat n'}} \cdot {\bf{\hat R}})} \right]\frac{{{\partial _t}T(t)}}{{4\pi {R^3}}}\vartheta ({\bf{r'}}){\ell _{qn}}({\bf{r'}})ds'} } \right|_{t = (j - i)\Delta t}}\\
\nonumber & + {\left. {{\beta _2}\frac{{{\rho _1}}}{{{\rho _2}}}\int_{{S_q}} {\left[ { - ({\bf{\hat n}} \cdot {\bf{\hat n'}}) + ({\bf{\hat n}} \cdot {\bf{\hat R}})({\bf{\hat n'}} \cdot {\bf{\hat R}})} \right]\frac{{\partial _t^3T(t)}}{{8\pi c_2^2R}}\vartheta ({\bf{r'}}){\ell _{qn}}({\bf{r'}})ds'} } \right|_{t = (j - i)\Delta t}}. 
\end{align}
The first and the third integrals on the right-hand side of Eq.~\eqref{eq22} and the first and the fourth integrals on the right-hand side of Eq.~\eqref{eq23} are ``smooth’’ and computed using a Gaussian quadrature rule~\cite{jin2010}. The third and the sixth integrals on the right-hand side of Eq.~\eqref{eq23} are weakly-singular and computed using the Duffy transformation~\cite{duffy1982quadrature}. The second and the fourth integrals on the right-hand side of Eq.~\eqref{eq22} and the second and the fifth integrals on the right-hand side of Eq.~\eqref{eq23} cancel out each other for ${\alpha _1} = {\alpha _2}$ and ${\beta _1}{\rho _2} = {\beta _2}{\rho _1}$, respectively.

\section{Numerical Results}\label{NR}
In this section, numerical results, which demonstrate the relationship between numerical errors and  interior resonance modes, are presented. TDPIE and TDCPIE are used to analyze acoustic scattering from a penetrable unit sphere that resides in an unbounded backgroud medium. It is assumed that the sphere is centered at the origin. The wave speed in the background medium and inside the sphere is ${c_1} = 300{\mathrm{\ m/s}}$ and ${c_2} = 200{\mathrm{\ m/s}}$, respectively. The ratio of the densities in these two media is ${\rho _1}/{\rho _2} = 1.5$. In all simulations, the excitation is a plane wave with velocity potential
\begin{align}
{\varphi ^{\mathrm{i}}}({\bf{r}},t) = {\varphi _0}G(t - {{\bf{\hat k}}^{\mathrm{i}}} \cdot {\bf{r}}/{c_1}), \label{eq24}
\end{align}
where ${\varphi _0}$ is the amplitude, ${{\bf{\hat k}}^{\mathrm{i}}}$ is the unit vector along the direction of propagation, and $G(t) = \cos [2\pi {f_0}(t - {t_{\mathrm{p}}})]\exp [ - {(t - {t_{\mathrm{p}}})^2}/(2{\sigma ^2})]$ is a modulated Gaussian pulse with center frequency ${f_0}$, time delay ${t_{\mathrm{p}}}$, and duration $\sigma $. The excitation parameters are selected as ${\varphi _0} = 1{\mathrm{\ }}{{\mathrm{m}}^2}{\mathrm{/s}}$, ${{\bf{\hat k}}^{\mathrm{i}}} = {\bf{\hat z}}$, ${f_0} = 120{\mathrm{\ Hz}}$, ${t_{\mathrm{p}}} = 10\sigma $, and $\sigma  = 3/(2\pi {f_{{\mathrm{bw}}}})$, where the effective bandwidth ${f_{{\mathrm{bw}}}} = 80{\mathrm{\ Hz}}$. Note that this  definition of $\sigma $ ensures that  $99.997\% $ of the wave energy is within the frequency band $ [{f_{\min }},{\mathrm{ }}{f_{\max }}]$ with ${f_{\min }} = {f_0} - {f_{{\mathrm{bw}}}}$ and ${f_{\max }} = {f_0} + {f_{{\mathrm{bw}}}}$~\cite{bagci2007fast}. Also, this specific selection of ${f_0} = 120{\mathrm{\ Hz}}$ and ${f_{{\mathrm{bw}}}} = 80{\mathrm{\ Hz}}$ ensures that the frequency of the lowest interior resonance mode (the first cavity mode of the Dirichlet problem), $150{\mathrm{\ Hz}}$, is within the frequency band $ [{f_{\min }},{\mathrm{ }}{f_{\max }}]$, i.e., this resonance mode could possibly be excited using the Gaussian pulse described above~\cite{zheng2015}. The time step size is chosen as $\Delta t = 1/(2\gamma f_{\mathrm{max}})$ with oversampling factor $\gamma $.  The half-width of the APSW interpolator used to construct $T(t)$ is ${N_{{\mathrm{hw}}}} = 7$. The surface of the sphere is discretized using ${N_{\mathrm{p}}} = 396$ curvilinear triangles with ${N_{\mathrm{n}}} = 6$ interpolation nodes on each triangle. For TDCPIE, the linear combination coefficients in Eqs.~\eqref{eq11} and~\eqref{eq12} are chosen as ${\alpha _1} = {\alpha _2} = {\beta _1} = 1$ and ${\beta _2} = 2/3$. LU decomposition is used to solve the matrix system in Eq.~\eqref{eq16} (at every time step) to ensure that the error in the matrix solution is at the machine precision level~\cite{anderson1999}. 

\subsection{Accuracy of TDPIE and TDCPIE} \label{ATT}
\begin{figure}
\centering
\includegraphics[width=0.68\textwidth]{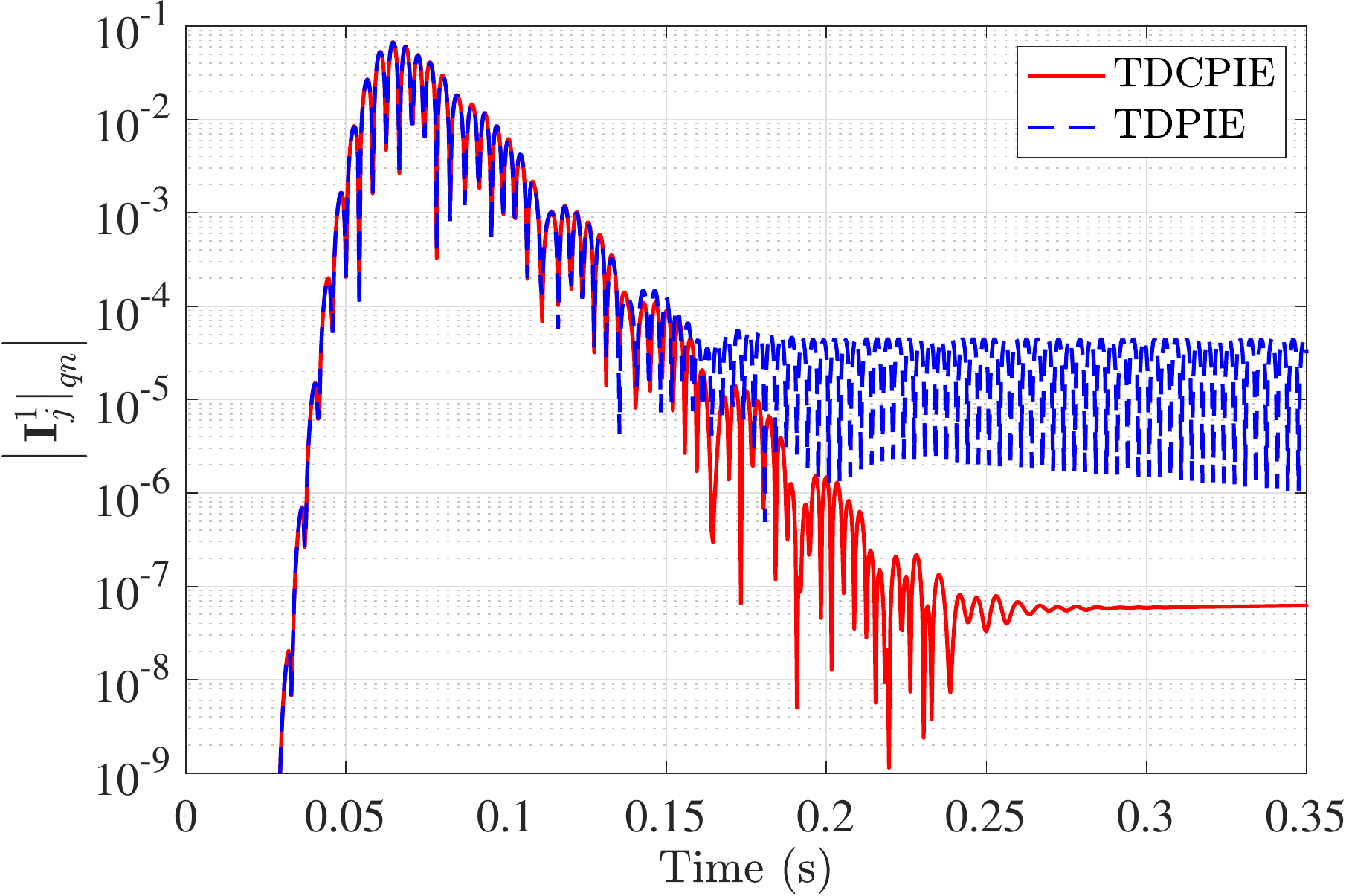}\\
\caption{Magnitude of the expansion coefficient ${\left. {{\bf{I}}_j^1} \right|_{qn}}$, $q = 3$, $n = 3$ (corresponding to ${{\bf{r}}_{qn}} = (0.78,0.58,0.23){\mathrm{\ m}}$), $j = 1, \ldots ,{N_{\mathrm{t}}}$ computed by solving TDPIE and TDCPIE.}
\label{fig2}
\end{figure}

\begin{figure}
\centering
\includegraphics[width=0.68\textwidth]{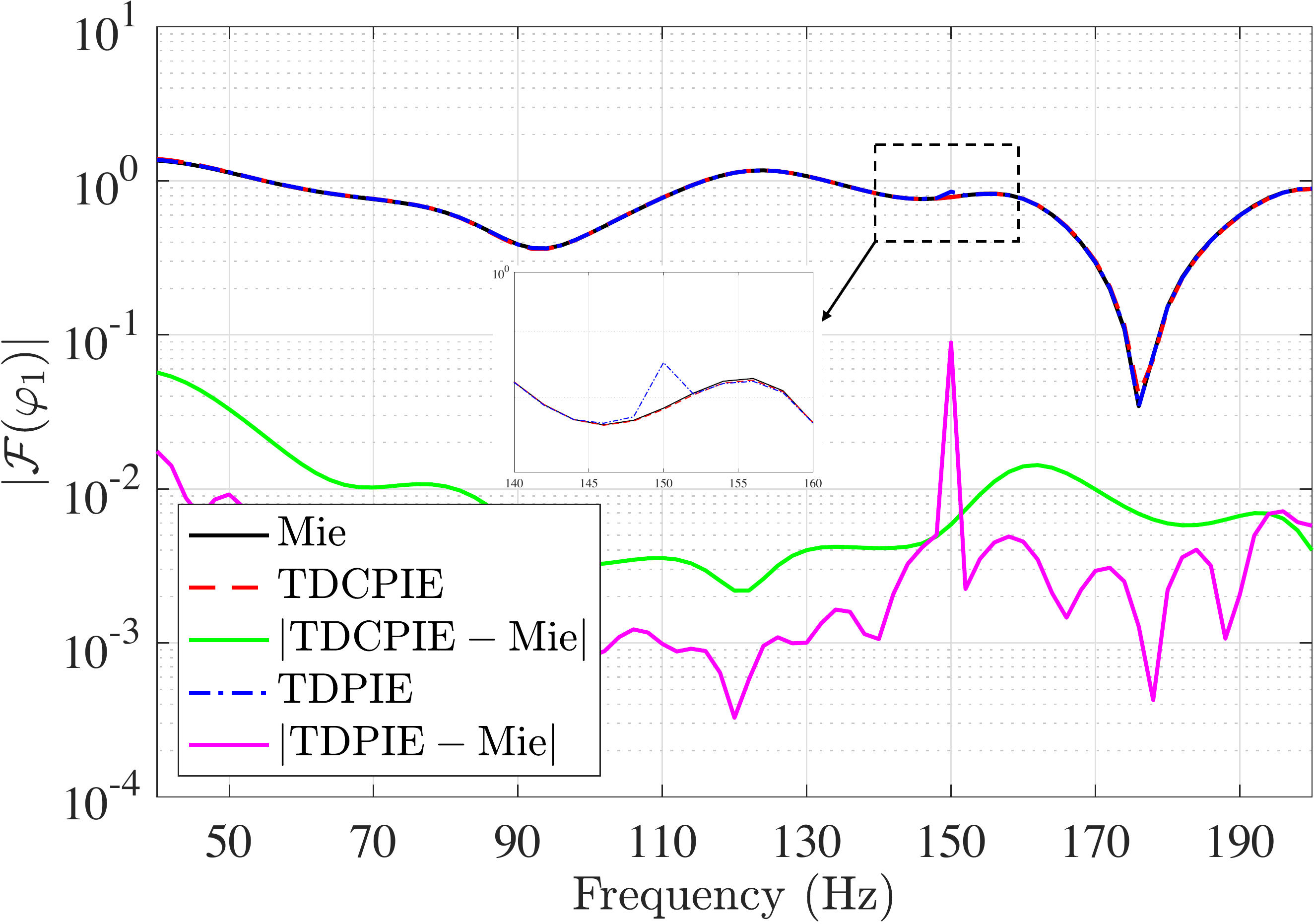}\\
\caption{Normalized Fourier transform of ${\varphi _1}({\bf{r}},t)$ computed after solving TDPIE and TDCPIE and the frequency-domain total velocity potential computed at ${{\bf{r}}} = (0.78,0.58,0.23{\mathrm{\ m}})$ using the Mie series.}
\label{fig3}
\end{figure}

For the first set of simulations, the oversampling factor is selected as $\gamma  = 6$ resulting in $\Delta t = 0.42{\mathrm{\ ms}}$. 16-point Gaussian and 9-point Gauss-Legendre quadrature rules~\cite{jin2010} are used to compute the two-dimensional (2D) surface integral with ``smooth’’ integrand and the one-dimensional (1D) line integral needed for the Duffy transformation~\cite{duffy1982quadrature}, respectively. Fig.~\ref{fig2} compares the magnitude of the expansion coefficient ${\left. {{\bf{I}}_j^1} \right|_{qn}}$, $q = 3$, $n = 3$ (corresponding to ${{\bf{r}}_{qn}} = (0.78,0.58,0.23)\,{\mathrm{m}}$), $j = 1, \ldots ,{N_{\mathrm{t}}}$ computed by solving TDPIE with that computed by solving TDCPIE. Clearly, the solution of TDPIE is corrupted by non-decaying oscillations while the solution of TDCPIE is free from any resonances. Fig.~\ref{fig3} shows the normalized Fourier transform of ${\varphi _1}({\bf{r}},t)$ [i.e., Fourier transform of  ${\varphi _1}({\bf{r}},t)$ divided by the Fourier transform of $G(t)$] computed after solving TDPIE and TDCPIE and compares that to the frequency-domain (time-harmonic) total velocity potential computed at ${{\bf{r}}} = (0.78,0.58,0.23)\,{\mathrm{m}}$ using the Mie series solution~\cite{turley2006acoustic}. Note that the normalization is required to ensure that the time-harmonic response (with equal excitation amplitude at each frequency) is obtained from the solutions of TDPIE and TDCPIE. Fig.~\ref{fig3} clearly shows that both simulations are accurate within the effective band of the excitation, except at $150{\mathrm{\ Hz}}$ where the solution of TDPIE is corrupted by the interior resonance mode. Fig.~\ref{fig3} also shows that TDPIE is more accurate than TDCPIE at other frequencies. This might be explained by the fact that that a second-kind surface integral equation (e.g., TDCPIE) is usually less accurate than its first-kind counterpart (e.g., TDPIE)~\cite{yan2012}.

\begin{figure}
\centering
\subfigure[]{\includegraphics[width=0.68\textwidth]{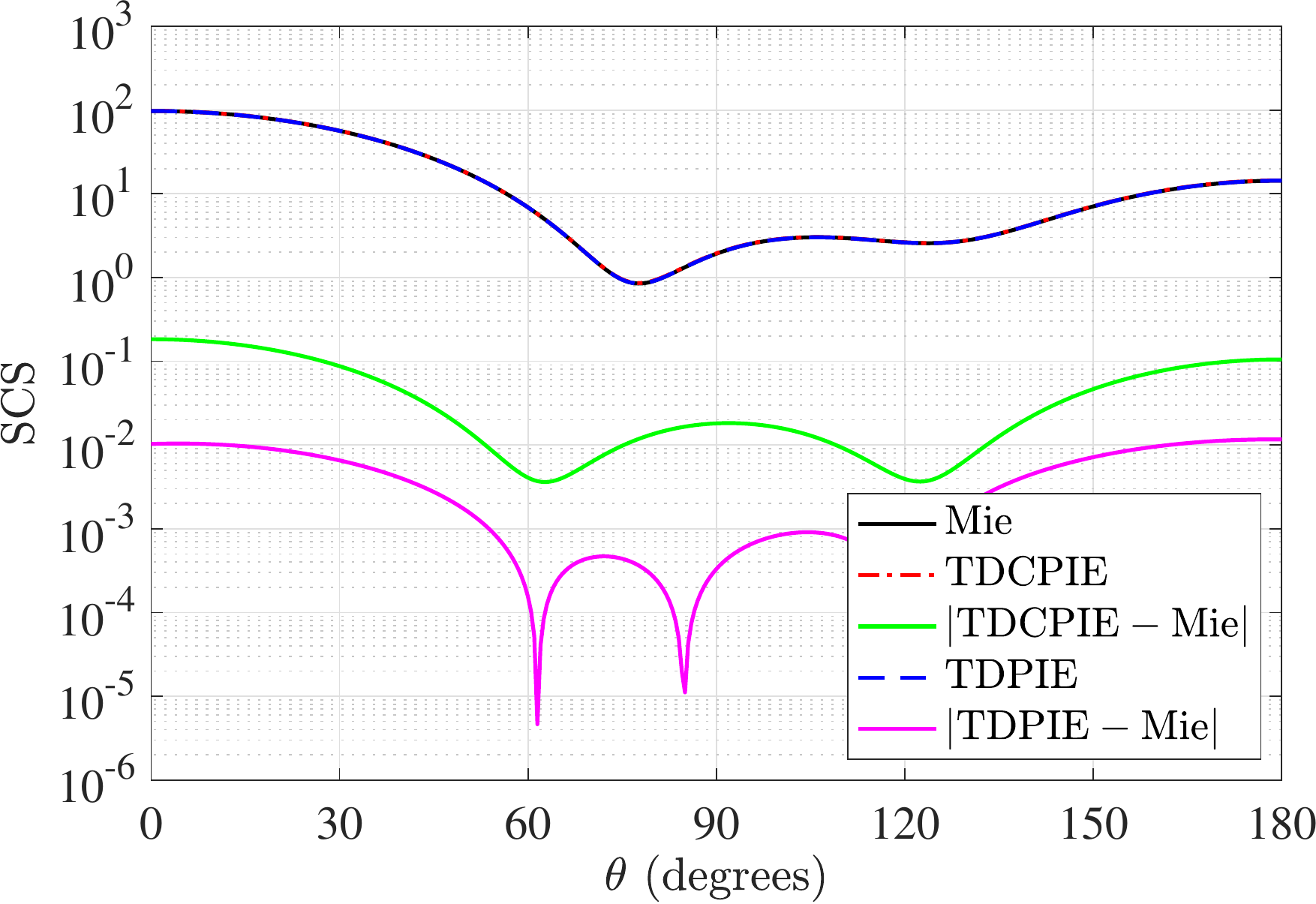}}\\
\subfigure[]{\includegraphics[width=0.68\textwidth]{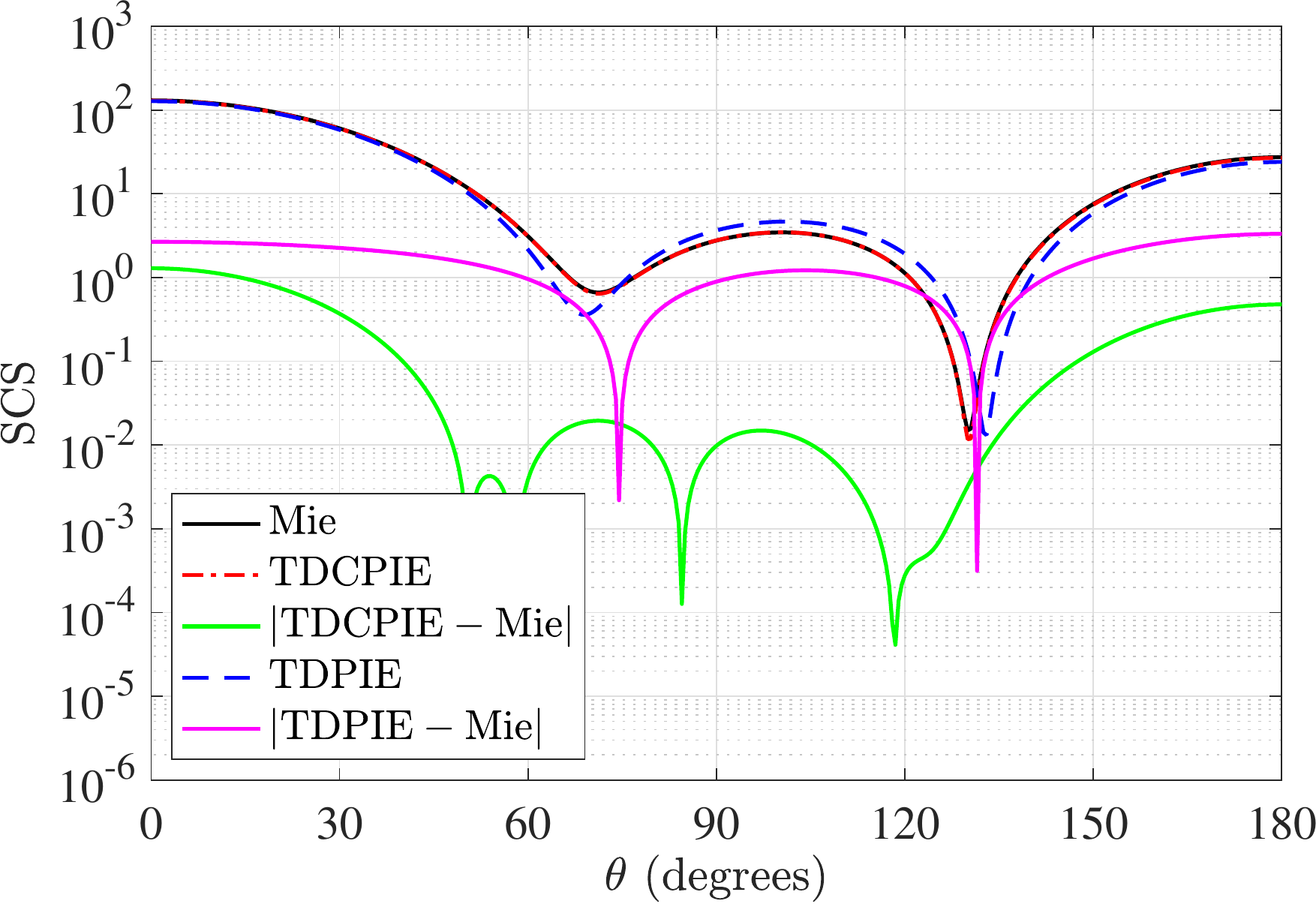}}\\
\caption{SCS computed using the Fourier transformed solutions of TDCPIE and TDPIE and the Mie series solution versus $\theta $ for $\phi  = {0^ \circ }$ at (a) $120{\mathrm{\ Hz}}$ and (b) $150{\mathrm{\ Hz}}$.}
\label{fig4}
\end{figure}

The effect of interior resonances on the scattered velocity potential is investigated by comparing the scattering cross section (SCS) of the sphere computed using the normalized Fourier transformed solutions of TDPIE and TDCPIE to SCS computed using the Mie series solution~\cite{turley2006acoustic}. Fig.~\ref{fig4}(a) and (b) plot SCS versus $\theta $ for $\phi  = {0^ \circ }$ at  $120\,{\mathrm{Hz}}$ and $150\,{\mathrm{Hz}}$, respectively. As shown in Fig.~\ref{fig4}(a), SCS computed using TDPIE and TDCPIE solutions at $120\,{\mathrm{Hz}}$ shows good agreement with the Mie results. On the other hand, as shown in Fig.~\ref{fig4}(b), the resonance mode at $150{\mathrm{\ Hz}}$ dramatically changes SCS computed using the TDPIE solution. 

\subsection{Effect of Numerical Integration Accuracy}\label{ENIA}
In this set of simulations, the effect of the computation accuracy of the integrals in Eqs.~\eqref{eq17} and~\eqref{eq18} on the amplitude of the interior resonance modes is investigated. Three sets of computation accuracy are considered by using different number of quadrature points to compute the 2D surface integral with ``smooth’’ integrand and the 1D line integral needed for the Duffy transformation~\cite{duffy1982quadrature}:
\begin{enumerate}[label=(\roman*)]
\item 16-point Gaussian and 9-point Gauss-Legendre quadrature rules
\item 7-point Gaussian and 5-point Gauss-Legendre quadrature rules
\item 4-point Gaussian and 3-point Gauss-Legendre quadrature rules
\end{enumerate}

\begin{figure}
\centering
\includegraphics[width=0.68\textwidth]{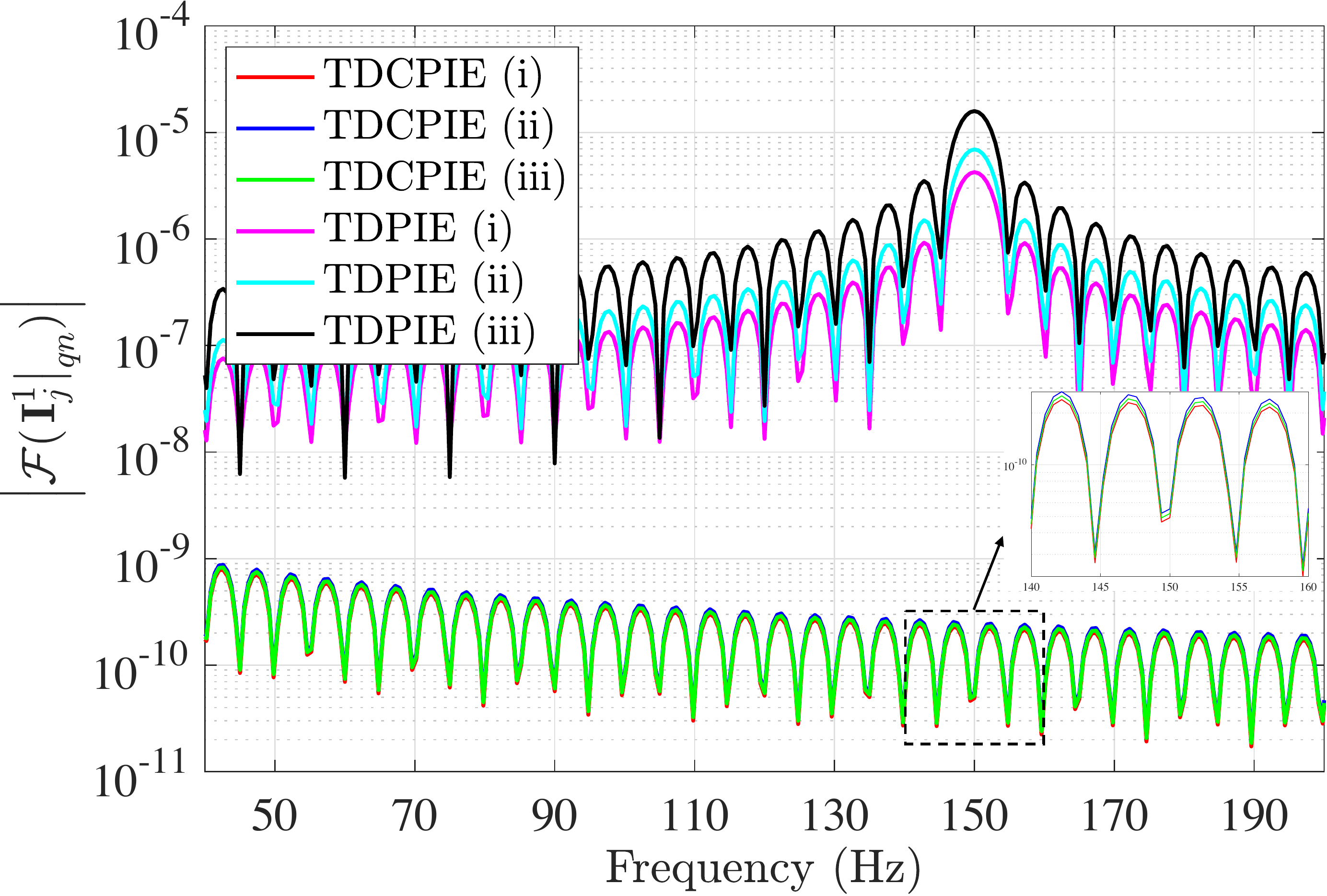}\\
\caption{Fourier transforms of ${\left. {{\bf{I}}_j^1} \right|_{qn}}$, $q = 3$, $n = 3$ (corresponding to ${{\bf{r}}_{qn}} = (0.78,0.58,0.23)\,{\mathrm{m}}$) computed after solving TDCPIE and TDPIE with the integration accuracy sets (i), (ii), and (iii) in the time range $t \in [0.8{\mathrm{\ s}},{\mathrm{ }}1{\mathrm{\ s}}]$.}
\label{fig5}
\end{figure}

\noindent In all simulations, the oversampling factor is selected as $\gamma  = 6$ resulting in $\Delta t = 0.42{\mathrm{\ ms}}$. To clearly identify the interior resonance mode, Fourier transforms of the late-time data of ${\left. {{\bf{I}}_j^1} \right|_{qn}}$, $q = 3$, $n = 3$ (corresponding to ${{\bf{r}}_{qn}} = (0.78,0.58,0.23)\,{\mathrm{m}}$) computed after solving TDCPIE and TDPIE with integration accuracy sets (i), (ii), and (iii) in the time range $t \in [0.8{\mathrm{\ s}},{\mathrm{ }}1{\mathrm{\ s}}]$ are plotted in Fig.~\ref{fig5}. The figure shows that the solutions of TDPIE with all three sets exhibit spurious interior resonance mode at $150{\mathrm{\ Hz}}$. However, as expected, no spurious resonance mode is present in the solutions of TDCPIE. Furthermore, the amplitude of the interior resonance mode observed in the solution of TDPIE with set (iii) is stronger than those with sets (i) and (ii). This is because set (iii) yields larger numerical errors than sets (i) and (ii). 

\begin{figure}
\centering
\subfigure[]{\includegraphics[width=0.68\textwidth]{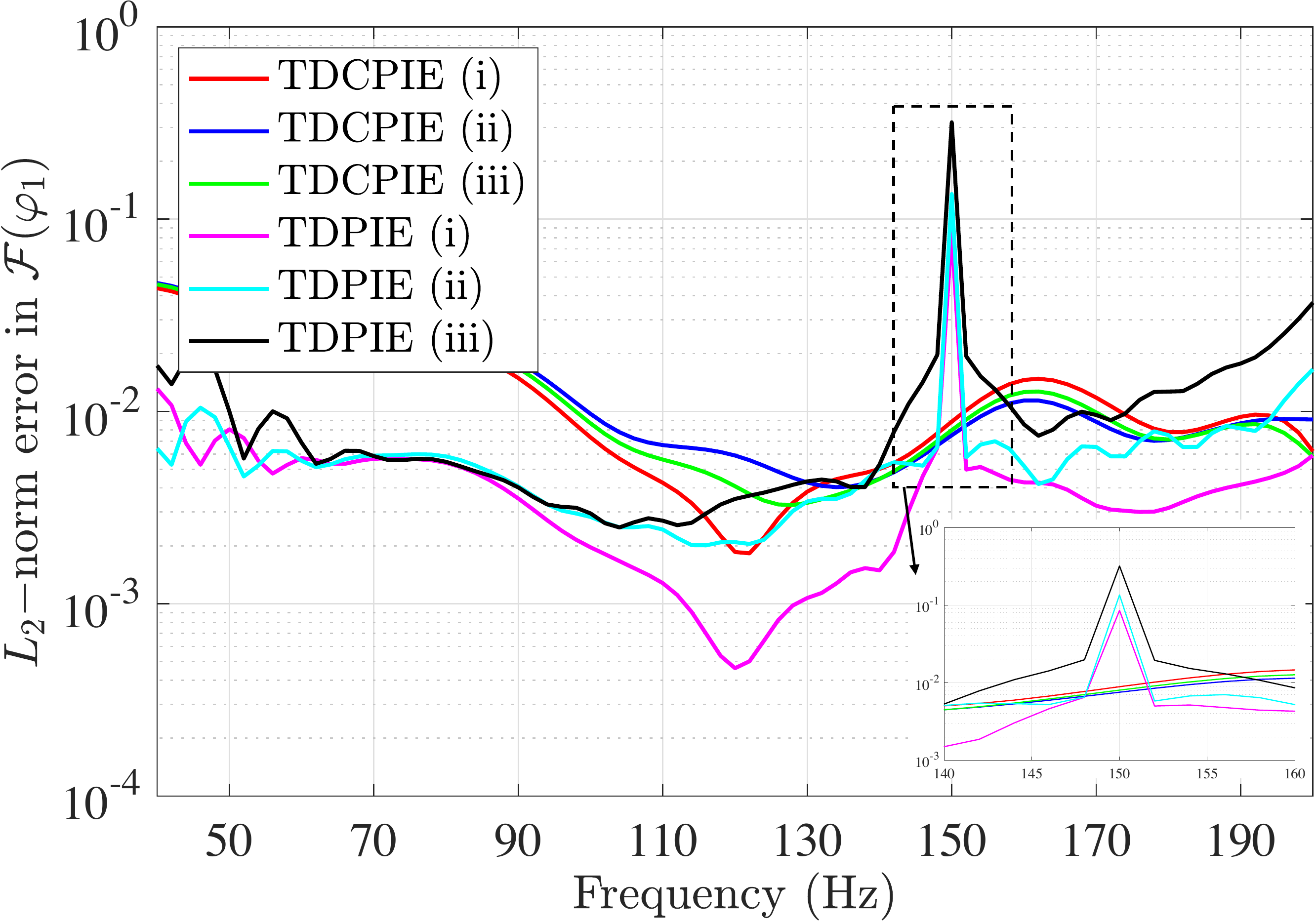}}\\
\subfigure[]{\includegraphics[width=0.68\textwidth]{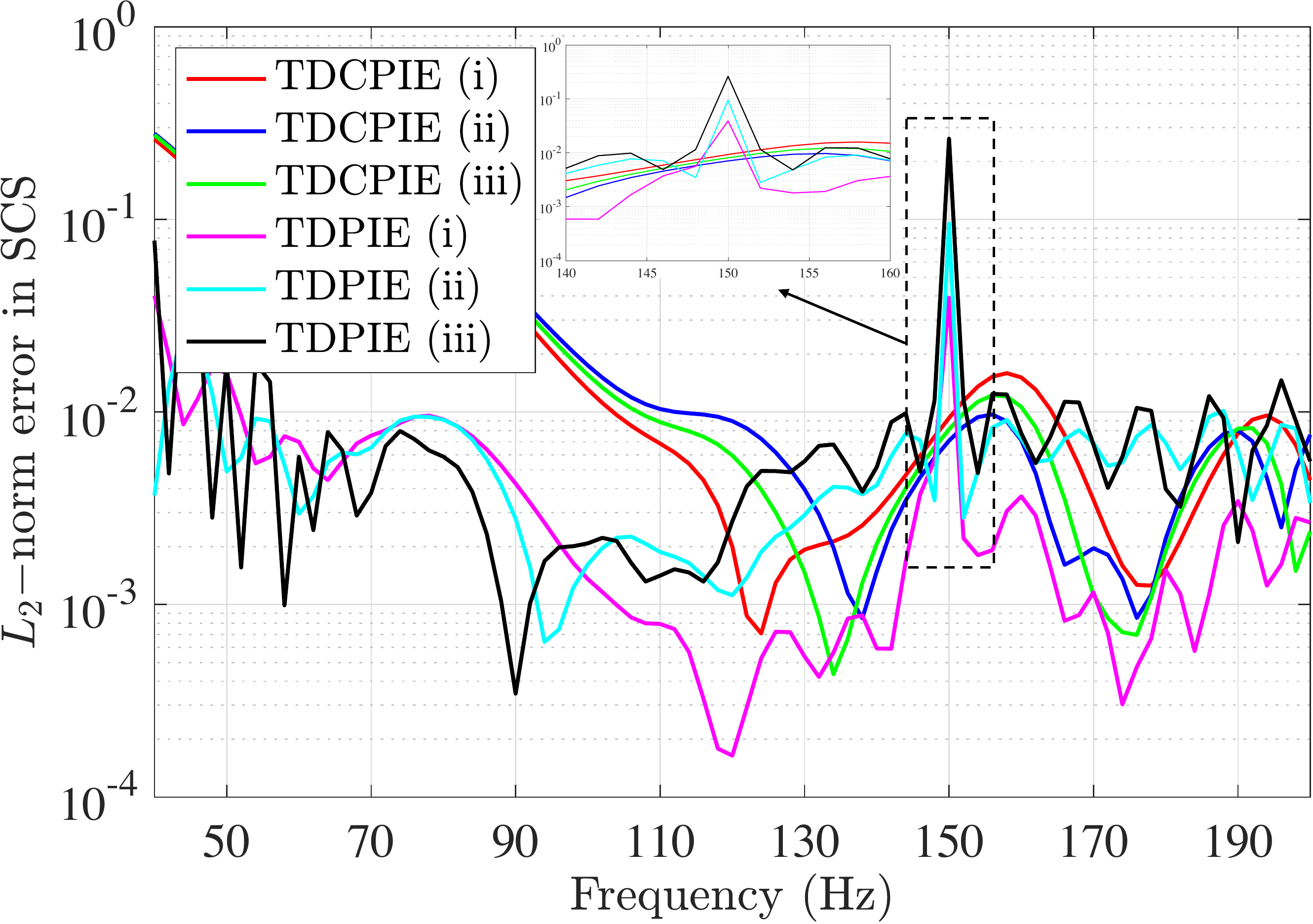}}\\
\caption{${L_2}$-norm error (a) in the normalized Fourier transform of ${\varphi _1}({\bf{r}},t)$ (at all the interpolation nodes on the sphere surface) and (b) in SCS for $\theta  = [{0^ \circ },{\mathrm{ }}{180^ \circ }]$ and $\phi  = {0^ \circ }$ (at $361$ samples points) computed after solving TDCPIE and TDPIE with the integration accuracy sets (i), (ii), and (iii).}
\label{fig6}
\end{figure}

To compare the accuracy of the solutions of TDCPIE and TDPIE with sets (i), (ii), and (iii), the ${L_2}$-norm error in the normalized Fourier transform of ${\varphi _1}({\bf{r}},t)$ (over all $2376$ interpolation nodes on the sphere surface) and the ${L_2}$-norm error in SCS (for $\theta  = [{0^ \circ },{\mathrm{ }}{180^ \circ }]$ with $361$ sample points and $\phi  = {0^ \circ }$) are plotted versus frequency in Fig.~\ref{fig6}(a) and (b), respectively. Note that the reference data used in the computation of the ${L_2}$-norm errors is obtained using the Mie series solution. Fig.~\ref{fig6} shows that the accuracy of the TDPIE solution in the vicinity of $150{\mathrm{\ Hz}}$ is significantly affected by the interior resonance mode. As expected, larger numerical errors increase the amplitude of the resonance mode. Fig.~\ref{fig6} also shows that the spurious interior resonance mode is not observed in the TDCPIE solution regardless of the integral computation accuracy.

\begin{figure}
\centering
\subfigure[]{\includegraphics[width=0.48\textwidth]{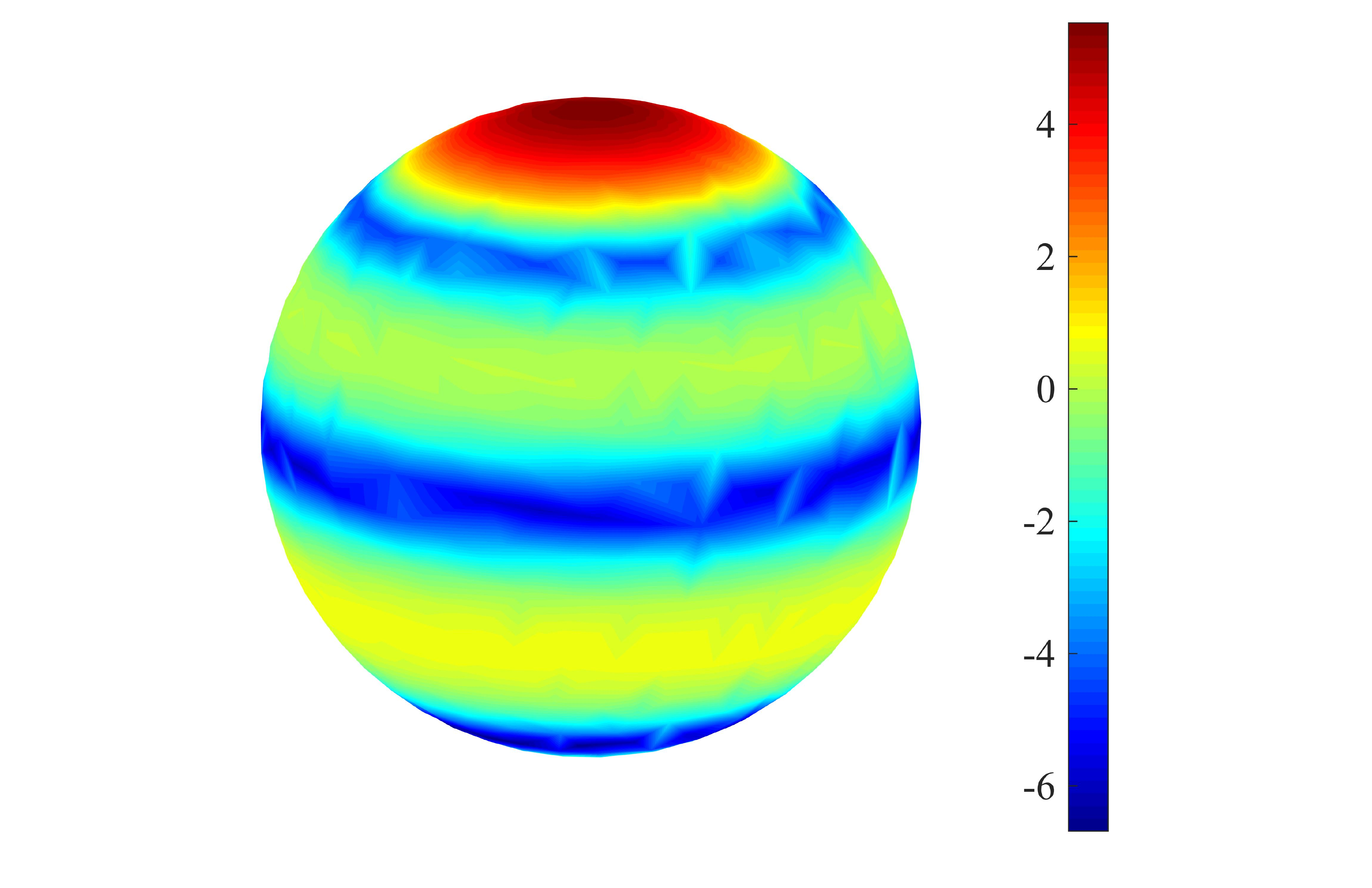}}
\subfigure[]{\includegraphics[width=0.48\textwidth]{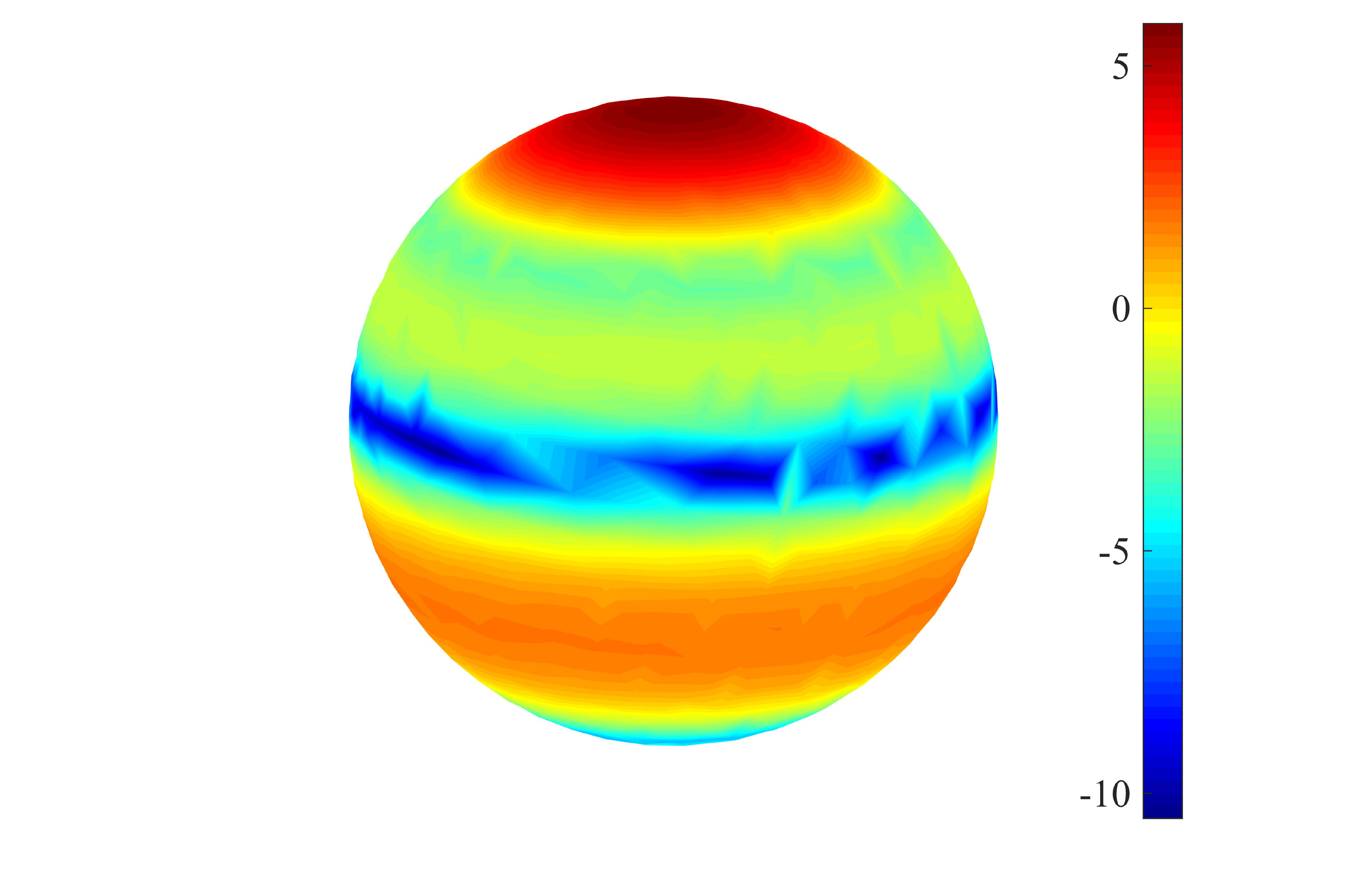}}\\
\subfigure[]{\includegraphics[width=0.48\textwidth]{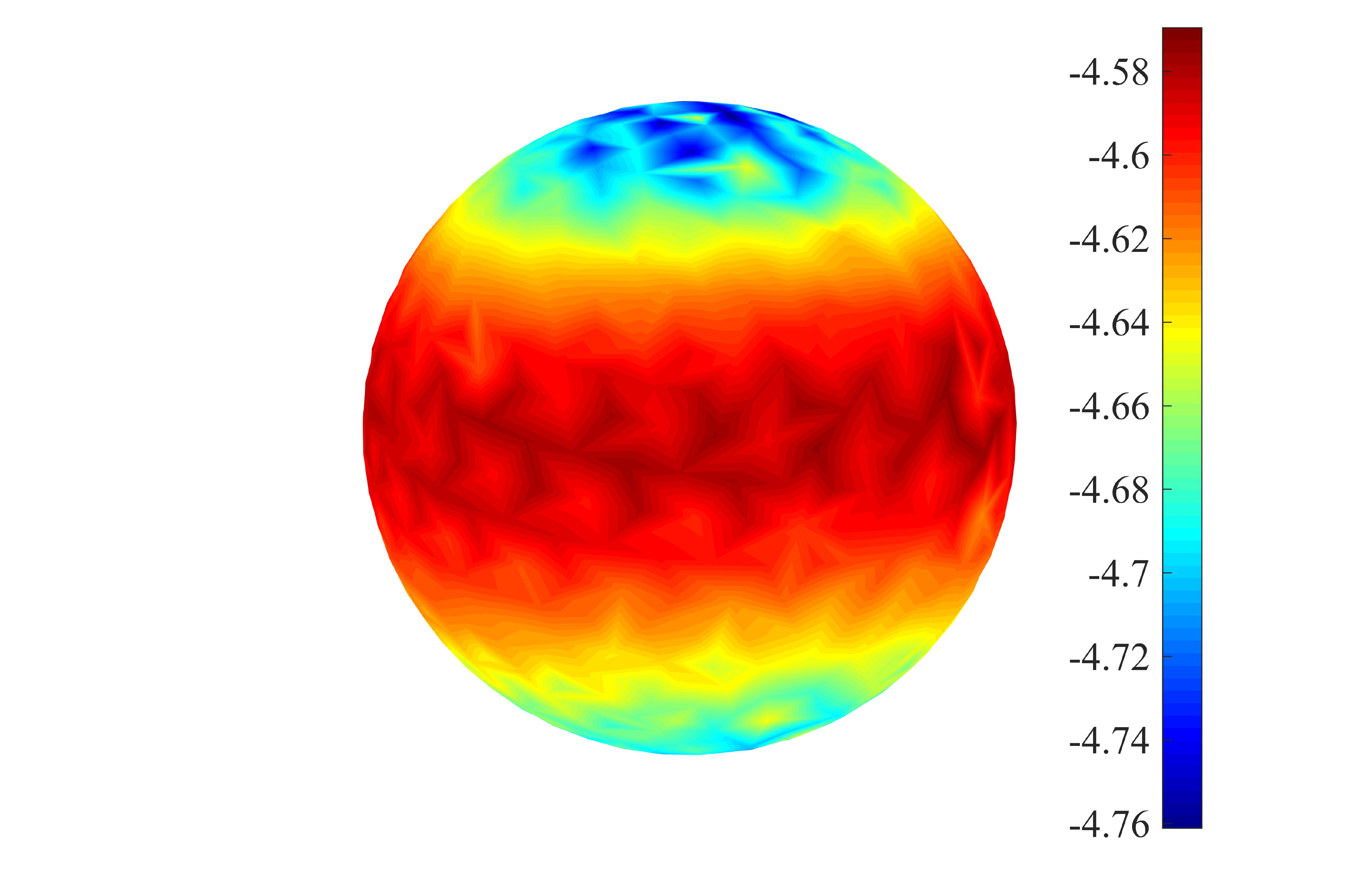}}\\
\caption{Patterns of the normalized Fourier transform of ${\varphi _1}({\bf{r}},t)$ over the sphere surface at $150{\mathrm{\ Hz}}$. (a) TDCPIE solution. (b) TDPIE solution. (c) Difference between TDCPIE and TDPIE solutions.}
\label{fig7}
\end{figure}

To visualize the interior resonance mode at $150{\mathrm{\ Hz}}$, the normalized Fourier transform of ${\varphi _1}({\bf{r}},t)$ (at all interpolation nodes on the sphere surface) computed after solving TDCPIE and TDPIE with set (iii) are presented in Fig.~\ref{fig7}(a) and (b), respectively. Fig.~\ref{fig7}(c) presents the difference in the normalized Fourier transform of ${\varphi _1}({\bf{r}},t)$ obtained from the TDCPIE and TDPIE solutions. As expected, the pattern of the difference follows the amplitude of the interior resonance mode~\cite{bagci2009}.

\subsection{Effect of Time Step Size}\label{ETSS}
\begin{figure}
\centering
\includegraphics[width=0.68\textwidth]{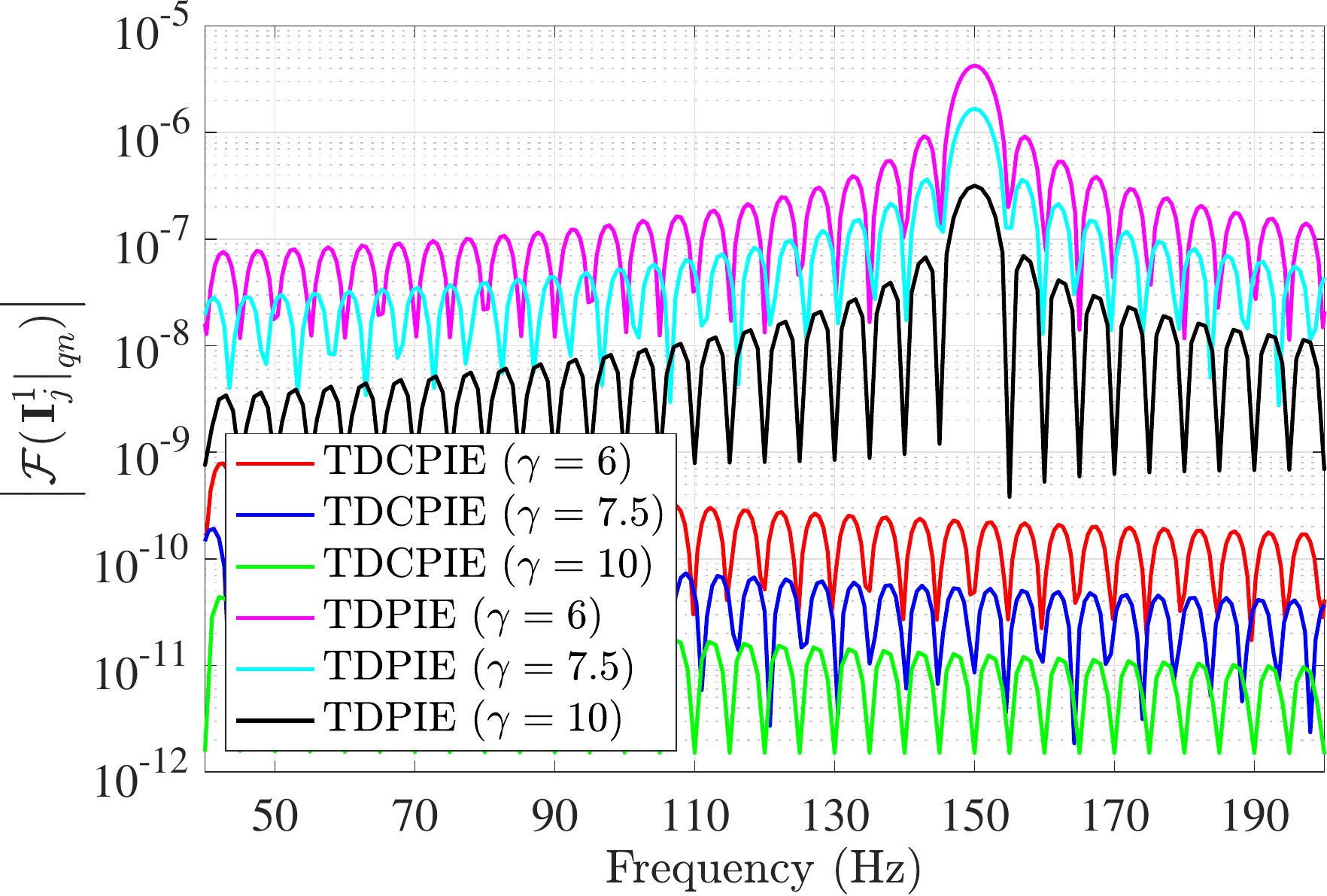}\\
\caption{Fourier transforms of ${\left. {{\bf{I}}_j^1} \right|_{qn}}$, $q = 3$, $n = 3$ (corresponding to  ${{\bf{r}}_{qn}} = (0.78,0.58,0.23)\,{\mathrm{m}}$) computed after solving TDCPIE and TDPIE with $\gamma  = 6$, $\gamma  = 7.5$, and $\gamma  = 10$ in the time range $t \in [0.8{\mathrm{\ s}},{\mathrm{ }}1{\mathrm{\ s}}]$.}
\label{fig8}
\end{figure}

\begin{figure}
\centering
\subfigure[]{\includegraphics[width=0.68\textwidth]{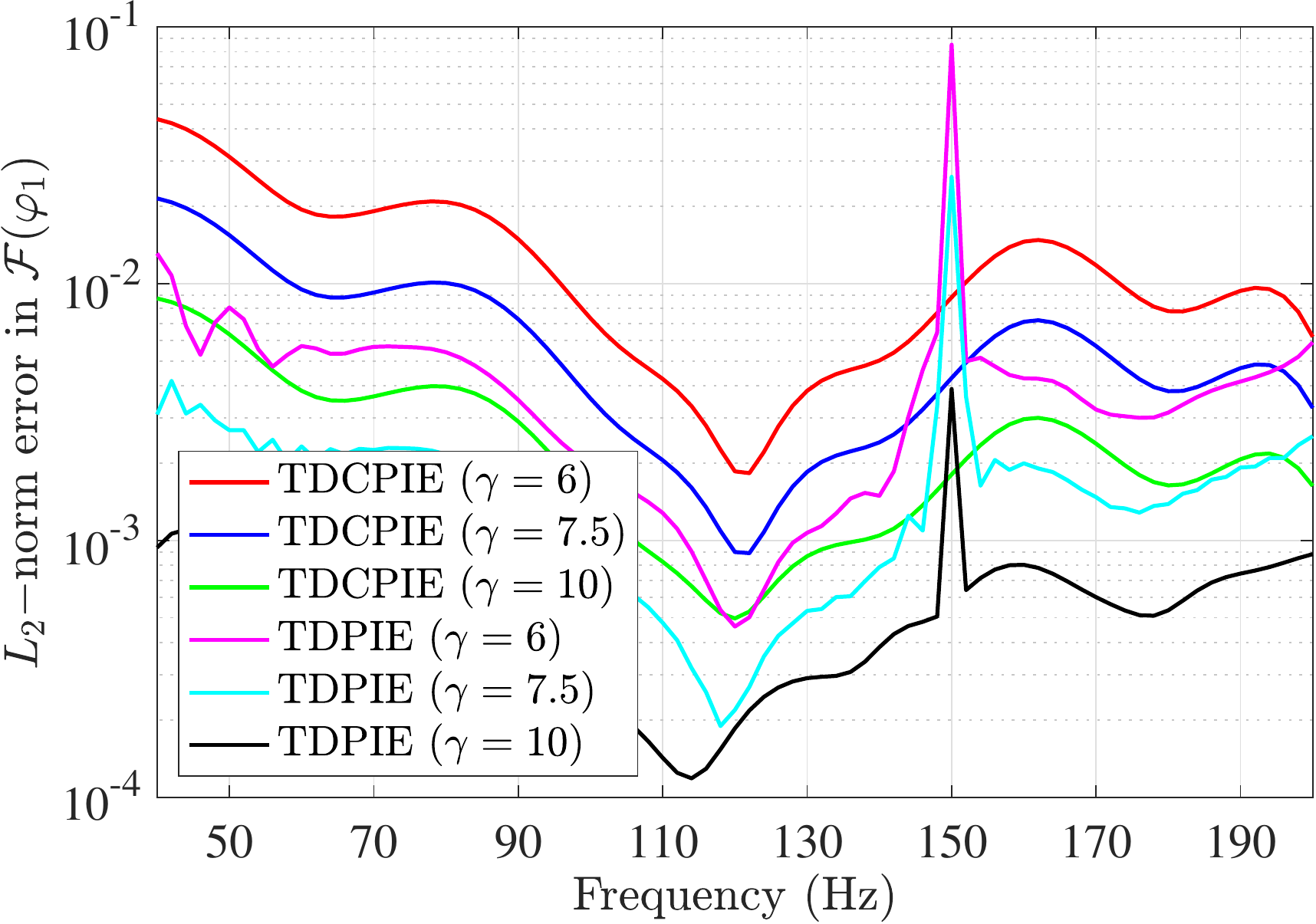}}\\
\subfigure[]{\includegraphics[width=0.68\textwidth]{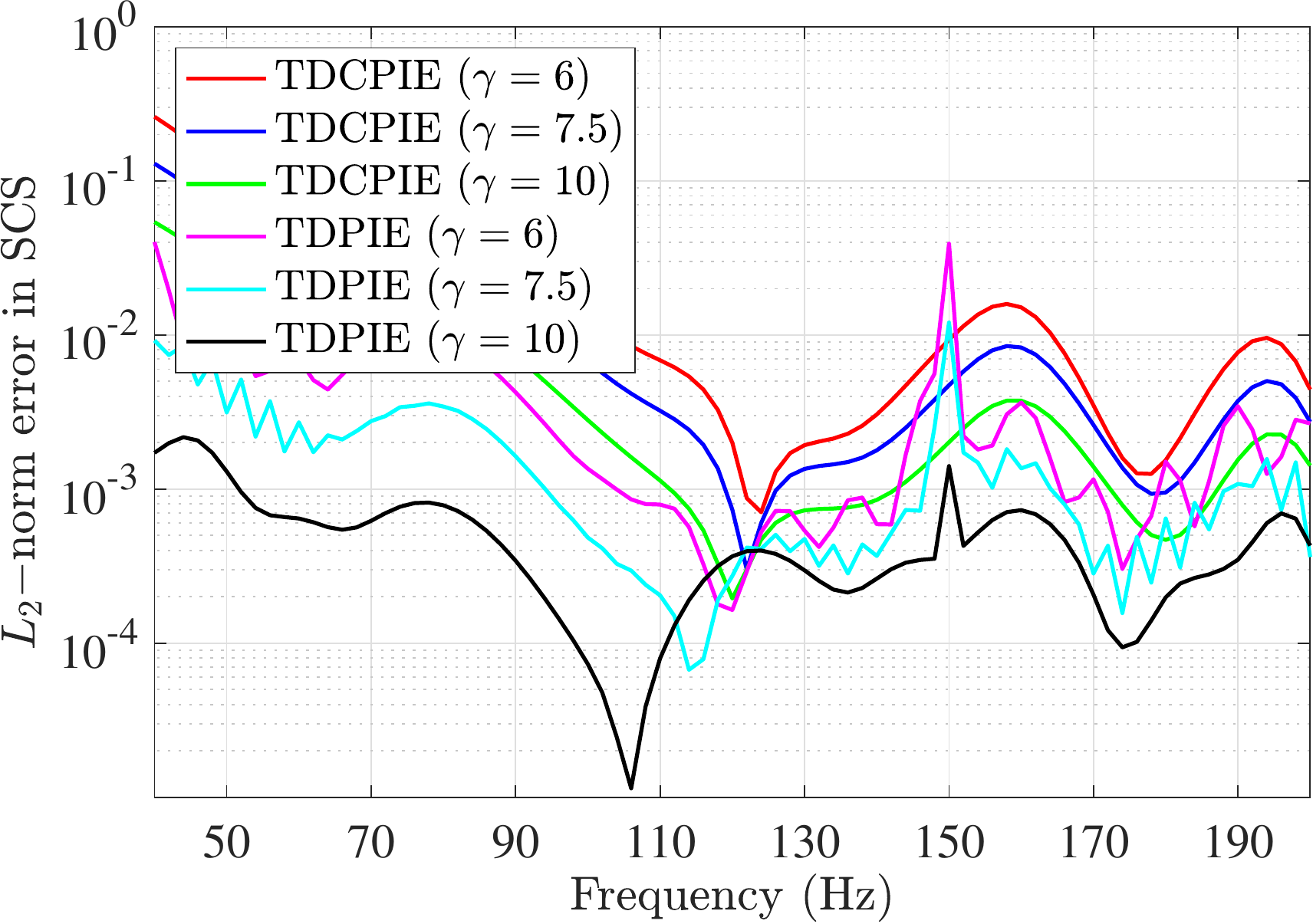}}\\
\caption{${L_2}$-norm error (a) in the normalized Fourier transform of ${\varphi _1}({\bf{r}},t)$ (at all the interpolation nodes on the sphere surface) and (b) in SCS for $\theta  = [{0^ \circ },{\mathrm{ }}{180^ \circ }]$ and $\phi  = {0^ \circ }$ (at  361 sample points) computed after solving TDCPIE and TDPIE with $\gamma  = 6$, $\gamma  = 7.5$, and $\gamma  = 10$.}
\label{fig9}
\end{figure}

In this set of simulations, the effect of time step size $\Delta t$ (i.e., temporal discretization density) on the amplitude of the interior resonance modes is demonstrated. Three different oversampling factors are used: $\gamma  \in \{ 6,7.5,10\} $ resulting in $\Delta t = \{ 0.42,0.33,0.25\} \,{\mathrm{ms}}$, respectively. In all simulations, the integration accuracy set (i) described in Section~\ref{ENIA} is used to ensure that the numerical error resulting from the computation of the space integrals in matrix elements is suppressed to be sufficiently small. Fig.~\ref{fig8} plots the Fourier transforms of ${\left. {{\bf{I}}_j^1} \right|_{qn}}$, $q = 3$, $n = 3$ (corresponding to ${{\bf{r}}_{qn}} = (0.78,0.58,0.23)\,{\mathrm{m}}$) computed after solving TDCPIE and TDPIE with $\gamma  = 6$, $\gamma  = 7.5$, and $\gamma  = 10$ in the time range $t \in [0.8{\mathrm{\ s}},{\mathrm{ }}1{\mathrm{\ s}}]$. As expected, interior resonance mode is observed in the TDPIE solution at $150{\mathrm{\ Hz}}$. Furthermore, the figure shows that using a larger $\gamma $ (or smaller $\Delta t$) reduces the amplitude of the resonance mode. As expected, no interior resonance mode is observed in the TDCPIE solution regardless of the value of $\gamma $ used. Fig.~\ref{fig9}(a) and (b) plot the ${L_2}$-norm error in the normalized Fourier transform of ${\varphi _1}({\bf{r}},t)$ (over all $2376$ interpolation nodes on the sphere surface) and the ${L_2}$-norm error in SCS (for $\theta  = [{0^ \circ },{\mathrm{ }}{180^ \circ }]$ at $361$ sample points and $\phi  = {0^ \circ }$) computed after solving TDCPIE and TDPIE with $\gamma  = 6$, $\gamma  = 7.5$, and $\gamma  = 10$, respectively. Note that the reference data used in the computation of the ${L_2}$-norm errors is obtained using the Mie series solution. Fig.~\ref{fig9} shows that interior resonance mode is observed in all of the TDPIE solutions with different $\gamma $ but its amplitude could be significantly suppressed by increasing $\gamma $. Another point to note here is that, even though the TDCPIE solution does not admit any interior resonance modes, it is usually less accurate than the TDPIE solution within the whole effective band of the excitation except in the vicinity of $150{\mathrm{\ Hz}}$.

\section{Conclusion}\label{Conclusion}
The interior resonance problem of TDPIE and TDCPIE that are formulated to analyze the time domain acoustic field interactions on penetrable scatterers is investigated. Numerical results demonstrate that the solution of TDPIE is corrupted by the spurious interior resonance modes that oscillate (without any decay) with the resonance frequencies of the acoustic cavity in the shape of the scatterer and has the density and the wave speed of the background medium. However, unlike the frequency-domain integral equations, the amplitude of these modes in the time domain can be suppressed by reducing the numerical error. On the other hand, the solution of TDCPIE, which is obtained by linearly combining TDPIE with its normal derivative, is free from spurious interior resonance modes.  It should be noted here that the weights of this linear combination are carefully selected to enable the numerical computation of the singular integrals.

\section*{Acknowledgements}
This publication is based upon work supported by the King Abdullah University of Science and Technology (KAUST) Office of Sponsored Research (OSR) under Award No 2019-CRG8-4056. The authors would like to thank the King Abdullah University of Science and Technology Supercomputing Laboratory (KSL) for providing the required computational resources.


\bibliographystyle{ieeetr} 
\bibliography{References}





\end{document}